\documentclass[12pt]{article}
\usepackage{amsmath}
\usepackage{amsfonts}

\usepackage{amsthm}
\usepackage{graphicx,psfrag,epsf}
\usepackage{enumerate}
\usepackage{framed}
\usepackage{algorithm}
\usepackage[noend]{algpseudocode}
\usepackage{flexisym}
\usepackage{subfloat}
\usepackage{subfig}

\usepackage{url} 
\newcommand{\blind}{1}
\DeclareGraphicsExtensions{.pdf,.png,.jpg}
\begin{document}

\def\spacingset#1{\renewcommand{\baselinestretch}%
{#1}\small\normalsize} \spacingset{1}


\if1\blind
{
  \title{\bf Validation of community robustness}
 
  \author{Annamaria Carissimo\thanks{
    All the authors equally contributed to the paper, \textit{corresponding author} {Luisa Cutillo: l.cutillo@sheffield.ac.uk}}\hspace{.2cm}\\
        Bioinformatics Core, TIGEM, Pozzuoli, Italy\\
    Luisa Cutillo \\
    University "Parthenope" of Naples and University of Sheffield\\
    and \\
   Italia Defeis\\
   Istituto per le Applicazioni del Calcolo "Mauro Picone", Naples, Italy}
 
  \maketitle
} \fi

\if0\blind
{
  \bigskip
  \bigskip
  \bigskip
  \begin{center}
    {\LARGE\bf Validation of community robustness}
    
\end{center}

  \medskip
} \fi

\bigskip

\begin{abstract}
The large amount of work on community detection and its applications leaves unaddressed one
important question: the statistical validation of the results. In this paper we present a methodology able to clearly detect if the community structure found by some algorithms is statistically significant or is a result of chance, merely due to edge positions in the network. Given a community detection method and a network of interest, our proposal examines the stability of the partition recovered against random perturbations of the original graph structure. To address this issue, we specify a perturbation strategy and a null model to build a set of procedures based on a special measure of clustering distance, namely Variation of Information, using tools set up for functional data analysis. The procedures determine whether the obtained clustering departs significantly from the null model. This strongly supports the robustness against perturbation of the algorithm used to identify the community structure. We show the results obtained with the proposed technique on simulated and real datasets.
\noindent%

\vfill

\end{abstract}

\section{Introduction}

\label{sec:intro}

Networks are mathematical representation of interactions among the components of a system and can be modeled by graphs. A graph G=(V,E) consists of a collection of vertices V, corresponding to the individual units of the observed system, and a collection of edges E, indicating some relation between pairs of vertices.

Graphs modeling real systems, i.e. social, biological, and technological networks, display non trivial topological features. Indeed they present the properties that define a complex network: big inhomogeneities, a broad degree distribution and  distribution of edges  locally inhomogeneous. In the study of complex networks, a network is said to have a community structure if the vertices can be divided
in $g$ groups, such that nodes belonging to the same group are densely connected and the number of edges between nodes of different groups is minimal.

The problem of community detection (graph partitioning) has been widely studied by researchers in a variety of fields, including statistics, physics, biology, social and computer science in the last 15 years. Finding communities within an arbitrary complex network can be a computationally difficult task. The number of communities, if any, within the network is typically unknown and the communities are often of unequal size and/or density. Despite these difficulties, however, several methods for community finding have been developed and employed with varying levels of success, see \cite{Coscia2011}, \cite{Fortunato2010}, \cite{Goldenberg2010}, \cite{Harenberg2014},
\cite{Kolacyzk2009} and \cite{Porter2009} for reviews.

Our work focusses on the problem of testing the robustness of the recovered partition of a given community detection method. In the following we provide a brief review of the state of the art of the literature addressing this problem. Although the huge work developed for community detection and its applications, the question of the significance of results still remains open.  Our proposal represents a first attempt to statistically define the robustness of a clustering and hence can not be directly compared to any of the following described methodologies. 

\paragraph{State of the art}
The modularity $Q$ of Newman and Girvan \cite{Newman2004} was the first attempt to give an answer to this question. It is defined as the fraction of the edges that fall within the given groups minus the expected such fraction if edges were distributed at random and is based on the idea that a random graph is not expected to have a cluster structure. However, as pointed out in  \cite{Fortunato2010} and \cite{Karrer2008}, there is an important limit. Precisely, networks with a strong community structure have high modularity, on the contrary high modularity does not imply networks with a community structure. Other authors, see \cite{Guimera2004}, \cite{Reichardt2007}, suggested the use of a z-score to compare the maximum modularity of a graph to the maximum attainable modularity in purely random graphs of the same size and expected degree sequence. The problem is that the distribution of the null model, though peaked, is not Gaussian, causing false positives and false negatives.

A different approach was developed in \cite{Massen2006}, where the authors studied how canonical ensembles of network partitions depend on temperature to assess the significance and nature of the community structure obtained by algorithms that optimize the modularity. In this case $-Q$ plays the role of energy, i.e. at temperature $T$, the statistical weight of a given partition in the ensemble is proportional to $\exp(Q/T)$. Typically, as the temperature increases, there is a transition from low entropy/high $Q$ partitions (significant cluster structure) to high entropy/low $Q$ (random partitions). If there is strong community structure, the transition is sharp. The peak is broader for networks with weaker community structure, as there are more reasonable alternative partitions with intermediate values of $Q$, and so the transition occurs over a broader range of temperature. They also introduced an order parameter to measure the similarity of the sampled partitions at a given temperature, i.e. whether there is just a single partition with high $Q$ or a number of competing partitions. Therefore, it is a useful tool to detect false positives. However the methodology is computationally onerous and cannot be easily generalized to other optimization methods.

In \cite{Bianconi2009} the authors introduced the notion of entropy of graph ensembles to assess the relevance of additional information about the nodes of a network using the information that comes from the topology of the network itself. The indicator of clustering significance $\Theta$ they introduce can also reveal statistical regularities that shed light on possible mechanisms underlying the network stability and formation. 

In \cite{Lancichinetti2011} the authors presented the Order Statistics Local Optimisation Method ($OSLOM$), a technique based on the local optimization of a fitness function, the C-score \cite{Lancichinetti2010}, expressing the statistical significance of a cluster with respect to random fluctuations. Given a subgraph $C$ in a graph $G$, the C-score measures the probability that the number of links connecting a node to nodes in $C$, where $C$ is embedded within a random graph, is higher than or equal to the value seen in the original graph $G$. This score permits to rank all the vertices external to $C$ (in increasing order of the C-score), having at least one connection with $C$, and to calculate its order statistic distribution $\Omega$. The minimum of $\Omega$ is the random variable whose cumulative is the score of the community $C$. To asses its significance a threshold parameter $P$ is fixed. The procedure is iterated to analyse the full network. The novelty of this approach is the local estimate of the significance, i.e. of single communities, not of partitions; on the contrary a serious limit is due to the lack of a data driven procedure to estimate $P$, indeed the authors fix its value to 0.1.

Recently,  \cite{Wilson2014}  proposed a testing based community detection procedure called Extraction of Statistically Significant Communities (ESSC). The ESSC procedure measures the statistical significance of connections between a single vertex and a set of vertices in undirected networks under a null distribution derived from the configuration model \cite{Bender1978}. Given an observed network $G_0$ with $n$ vertices and a vertex set $B$, they introduce the statistics $\hat{d}(u:B)$, measuring the number of edges between a vertex $u$ and $B$ in the random model $\hat{G}$, and show that $\hat{d_n}(u:B)$ is approximately binomial as $n\rightarrow \infty$ in the total variation distance between two probability mass functions. This permits to obtain the p-values of the null distribution using the binomial approximation and gives origin to an iterative deterministic procedure that recovers robust communities. The technique has some similarities with OSLOM, indeed both are extraction methods and use the configuration model as reference distribution, but differentiates because the probabilities have a closed form.

Another group of techniques was proposed in \cite{Gfeller2005},\cite{Karrer2008}, \cite{Rosvall2010}, and their conceiving was completely different from previous described methodologies. Indeed they introduce a stochastic component in the network by perturbing the graph structure, measure the effect of the perturbation and compare it with the corresponding value for a null model graph. The basic idea is that a significant partition should not be altered by small modifications, as long as the modification is not too extensive. An interesting feature of these methods is their independence from the community detection technique adopted.

In this paper we present a methodology able to clearly detect if the community structure found by some algorithms is statistically significant or is a result of chance, merely due to edge positions in the network. Given a community detection method and a network of interest, our proposal examines the stability of the partition recovered against random perturbations of the original graph structure.  To address this issue, following ideas from \cite{Karrer2008}, we specify a perturbation strategy and a null model to build some procedures based on Variation of Information as stability measure. Given a this measure we address the question of evaluating its the significance. This permits to build the Variation of Information curve as a function of the perturbation percentage and to compare it with the corresponding null model curve using analysis tools set up for functional data analysis.

The rest of the paper is organised as follows. In Section \ref{VI} we introduce the proposed procedures based on Variation of Information and the functional data analysis techniques, including their detailed description. Section \ref{Results} shows the results achieved applying our methodology on simulated and real datasets. Conclusions and ideas for future research are drawn in Section \ref{Discussion}.

\section{Variation of Information hypothesis testing} \label{VI}

Variation of Information (VI) is an information theoretic criterion for comparing two partitions, or clusterings, of the same data set \cite{Meila2007}. It is a metric and measures the amount of information lost and gained in changing from clustering $\mathcal{C}$ to clustering $\mathcal{C'}$. The criterion makes no assumptions about how the clusterings were generated and applies to both soft and hard clusterings.

Given a dataset $D$ and two clusterings $\mathcal{C}$ and $\mathcal{C'}$ of $D$, with $K$ and $K'$ non empty clusters, respectively, VI is defined as
\begin{equation}
VI\left(\mathcal{C,C'}\right)=H\left(\mathcal{C}\right)+H\left(\mathcal{C'}\right)-2I\left(\mathcal{C,C'}\right)
\end{equation}
where $H\left(\mathcal{C}\right)$ is the entropy associated with clustering $\mathcal{C}$
\begin{equation}
H\left(\mathcal{C}\right)=-\sum_{k=1}^K P(k)\log P(k),
\end{equation}
and $I\left(\mathcal{C,C'}\right)$ is the mutual information between $\mathcal{C}$ and $\mathcal{C'}$, i.e the information that one clustering has about the other
\begin{equation}
I\left(\mathcal{C,C'}\right)=\sum_{k=1}^K \sum_{k'=1}^{K'} P\left(k,k'\right)\log \frac{P\left( k,k'\right)}{P\left(k\right)P\left(k'\right)}.
\end{equation}
$P\left(k\right)$ is the probability of a point being in cluster $C_k$ and $P\left(k,k'\right)$ is the probability that a point belongs to $C_k$ in clustering $\mathcal{C}$ and to $C_{k'}$ in $\mathcal{C'}$.

Another equivalent expression for VI is
\begin{equation}
VI\left(\mathcal{C,C'}\right)=H\left(\mathcal{C}|\mathcal{C'}\right)+H\left(\mathcal{C'}|\mathcal{C}\right).
\end{equation}
The first term measures the amount of information about $\mathcal{C}$ that we loose, while the second measures the amount of information about $\mathcal{C'}$ that we gain, when going from clustering $\mathcal{C}$ to clustering $\mathcal{C'}$.

VI metric is the basis of the hypothesis testing procedures we
propose to establish the statistical significance of a recovered
community structure in a complex network. 
Our original idea is to generate two different curves based on the VI measure and to
statistically test their difference. The first curve $VIc$ is
obtained computing VI between the partition of our original network
and the partition of different perturbed version of our original
network. The second curve $VIc_{random}$ is obtained computing
VI between the partition of a null random network and the partition
of different perturbed version of such null network. The comparison between the two VI curves turns the question about 
the significance of the retrieved community structure into the study 
of the stability/robustness of the recovered partition against perturbations. 
We expect that it must be robust to small perturbations, because if ``small changes" 
in the network imply a completely different partition of the data, it means that the found communities are not trustworthy, and this cannot be due to the failure of the chosen algorithm for the community detection. Indeed the proposed testing procedure is independent from the clustering algorithm and it is easy to check if such a behaviour is due to it.

To understand well this point we must consider the behaviour of the VI curve for networks having a real community structure and those having a very poor community structure. In the first case the VI curve starts at 0, when the perturbation level $p$ is 0$\%$ (unperturbed graph), rises rapidly (perturbation level between 0$\%$ and 40$\%$), then levels off when $50\%<p<100\%$; in the second case the VI curve immediately grows up to a certain value and levels off that value, meaning that whatever partition has been found, at each level of perturbation, it has the same robustness of a random graph.  Obviously the set up of a testing procedure is more necessary for all cases where the community structure is moderate or weak and the behaviour of the VI curve could be similar to that of a random graph.

The choice of the null random network is more delicate, because we expect that it has the same structure of our original graph but with completely random edges. This is why our choice relapses on the Configuration Model \cite{Bender1978} associated with the degree sequence of the observed graph $\mathbf{d}=\{d(1),\dots,d(n)\}$ with vertex set $V=\{1,\dots,n\}$, i.e. CM($\mathbf{d}$). The CM($\mathbf{d}$) is a probability measure on the family of multi-graphs with vertex set V and degree sequence $\mathbf{d}$ that reflects, within the constraints of the degree sequence, a random assignment of edges between vertices. The generative form is simple: one can simply cut all the edges in the network, so every node still retains its degree by the number of half-edges or stubs emanating from it. The result will be an even number of half-edges. To create new networks with the same degree, one simply needs to randomly pair all the half-edges, creating the new edges in the network. The Configuration Model generates every possible graph with the given degree distribution with equal probability.
Note that it naturally creates networks with multiple edges between nodes and self-connections between nodes. If such networks are unacceptable, one can reject those samples and try the algorithm again, repeating until one obtains a network without multiple or self-connections. The perturbation strategy adopted is described in Section \ref{permute}. The basic
steps of our method are the following:

\begin{framed}\center{\textbf{Overall Procedure}}\label{procedure}

\begin{enumerate}
\item \label{first} find a partition $\mathcal{C}$ of the given network, with vertex set $V=\{1,\dots,n\}$ and degree sequence $\mathbf{d}=\{d(1),\dots,d(n)\}$, by some chosen method $M$;
\item \label{second} given a perturbation level $p\in \left[0,1\right]$, perturb the network shuffling its edges by the percentage $p$ preserving the original graphs degree distribution;
\item \label{third} using the same method $M$ find a partition $\mathcal{C'}$ for the perturbed network and compute the $VI(\mathcal{C},\mathcal{C'})$ to compare $\mathcal{C}$ and $\mathcal{C'}$ ;
\item \label{fourth} repeat steps \ref{second}-\ref{third} at different perturbation levels $p \in \left[0,1\right]$ so to obtain the $VIc$ curve. ( Note that $p=0$ corresponds to the original unperturbed graph, $p=1$ corresponds to the maximal perturbation level (random graph));
\item \label{fifth} repeat steps \ref{first}-\ref{fourth} starting in step \ref{first} from the CM($\mathbf{d}$) (null model) so to get the $VIc_{random}$ curve;
\item \label{six} compare the $VIc$ and $VIc_{random}$ curves as functions of $p$, statistically testing {\it ``the difference''} between the two curves.
\end{enumerate}
\end{framed}

Note that the variation of $p$ from 0 to 1 induces an intrinsic order to the data structure as in temporal data and can be treated as time point. Moreover, as it will be described in Section \ref{permute}, we generate many perturbed  graphs (i.e. $10$) for each different level of $p$ and these are considered as replicates per time points in our strategy.\\

Step \ref{six}  of the above procedure is achieved by a functional data analysis approaches  aiming to test if the two groups of curves represent ``the same process" or ``different processes". The testing procedure we rely on is  based on a tool set up for time course microarray data, namely Gaussian Process (GP) regression \cite{Kalaitzis2011}. 
Aim of the GP regression in the context of gene expression data is to identify differentially expressed genes in a one-sample time course microarray experiment, i.e. to detect if the profile has a significant underlying signal or the observations are just random fluctuations. In this case we reformulate the testing problem working on $\mathrm{\log_2} \frac{VIc}{VIc_{random}}$, as described in Section \ref{sec:GP}.

In order to show that our approach is robust with respect to the testing procedure used in Step \ref{six}, we also display the overall results achieved when using other two approaches in Step \ref{six}, described respectively in Section \ref{sec:FPC} and Section \ref{sec:IFT}. Indeed, we can look at the two measured $VI$ curves as independent realisations of two underlying processes say $X_1$ and $X_2$ observed with noise on a finite grid of points $p \in [0,1]$ and to test the null hypothesis 
\begin{equation}\label{H0_fadtest}
H_0: \mathrm{X_1 \buildrel d \over{=} X_2}
\end{equation} 
versus the alternative hypothesis 
\begin{equation*}
H_1: \mathrm{X_1\buildrel d \over{\neq }X_2}
\end{equation*} 
where $\buildrel d \over{=}$ means that the processes on either side have the same distribution. 

Then, as described in Section \ref{sec:FPC}, taking advantage of the Karhunen- Lo\`{e}ve expansion we explore the methodology developed in \cite{Pomann2016} based on Functional Principal Components Analysis (FPCA) to test (\ref{H0_fadtest}).

On the contrary, the approach described in Section \ref{sec:IFT} addresses a domain-selective inferential procedure, providing an interval-wise non parametric functional testing \cite{Pini2016}, able not only to assess (\ref{H0_fadtest}), but also to point out specific differences.

We will briefly describe GP regression, FPCA and interval-wise functional testing in the following sections. We would like to point out that our overall procedure provides a workflow to validate a community structure under different perspectives that can be investigated in dependence of the specific real problem dealt with. The description of the three different testing procedure is functional to the understanding of our overall procedure and in particular how we exploit the theory underling each single methodology to compare the  curves $VIc$ and
$VIc_{random}$.  Hence we will summarise the three testing procedures to highlight the key connection to our testing problem. We choose to provide a review of each single methodology to provide awareness of the differences between the three procedures and their link to our testing problem. Even if addressing the same problem they are not equivalent. We refer to the original papers for any theoretical property of such testing procedures, including type $I/II$ error study. We want to stress that the original contribution of our proposal is summarised in the six steps procedure depicted in the above frame.

\subsection{GP regression} \label{sec:GP}
In this section we briefly summarise the methodology proposed in \cite{Kalaitzis2011}, where the authors present an approach to estimate the continuous trajectory of gene expression time-series from microarray through GP regression. 

Briefly we recall that a gaussian process is the natural generalisation of a multivariate Gaussian distribution to a Gaussian distribution over a specific family of functions. More precisely, as defined in \cite{gpbook}, a gaussian process is a \textit{collection of random variables, any finite number of which have a joint Gaussian distribution} and is completely specified by its mean function and its covariance function. If we define the mean function $m(x)$ and the covariance function $k(x,x')$ of a real process $f(x)$ as:
\begin{align*}
m(x) &= E[f(x)],\\
k(x,x\textprime) &= E[(f(x) - m(x)(f(x\textprime) - m(x\textprime))]
\end{align*}

then we can write the GP as
\begin{equation}
f(x) \sim \mathcal{GP}(m(x), k(x,x\textprime)).
\end{equation}
The random variables $\mathbf{f}=\left(f\left(X_1\right),\dots,f\left(X_n\right)\right)^T$ represent the value of the function $f(x)$ at time locations $\left(X_i\right)_{i=1,\dots,n}$, being $f(x)$ the true trajectory/profile of the gene. Assuming $f(x)=\Phi(x)^T\mathbf{w}$, where $\Phi(x)$ are projection basis functions, with prior $\mathbf{w} \sim N(\mathbf{0},\sigma_{\mathbf{w}}^2\mathbf{I})$, we have
\begin{eqnarray}
&&E[f(x)]=\Phi(x)^TE[\mathbf{w}]=0 \label{GPbayes_mean}\\
&&E[f(x)f(x)\textprime]=\sigma_{\mathbf{w}}^2\Phi(x)^T\Phi(x) \label{GPbayes_cov}\\
&&f(x) \sim \mathcal{GP}(0,\sigma_{\mathbf{w}}^2\Phi(x)^T\Phi(x)).
\end{eqnarray} 
Since observations are noisy, i.e. $\mathbf{Y}=\mathbf{\Phi w}+\pmb{\varepsilon}$, with $\mathbf{\Phi}=(\Phi(X_1)^T,\dots,\Phi(X_n)^T)$, assuming that the noise $\pmb{\varepsilon} \sim N(\mathbf{0},\sigma_n^2\mathbf{I})$ and using 
Eq. (\ref{GPbayes_mean})-(\ref{GPbayes_cov}), the marginal likelihood 
\begin{eqnarray*}
p(\mathbf{y} | \mathbf{x})=\int p(\mathbf{y} | \mathbf{x},\mathbf{w}) p(\mathbf{w})d\mathbf{w}
\end{eqnarray*}
becomes
\begin{eqnarray}
p(\mathbf{y} | \mathbf{x})=\frac{1}{(2\pi)^{n/2}\left|\mathbf{K_y}\right|^{1/2}}\mathrm{exp}\left(-\frac{1}{2}\mathbf{y}^t\mathbf{K_y}^{-1}\mathbf{y}\right) \label{GPmarginal}
\end{eqnarray}
with $\mathbf{K_y}=\sigma_{\mathbf{w}}^2\mathbf{\Phi}\mathbf{\Phi}^T+\sigma_n^2\mathbf{I}$.

In this framework the hypothesis testing problem can be reformulated, over the perturbation interval $[0,1]$,  as:
\begin{eqnarray*}
H_0 : \mathrm{\log_2} \frac{VIc(x)}{VIc_{random}(x)} \sim \mathcal{GP}(0, k(x,x\textprime)) \label{H0_GP}
\end{eqnarray*}
against
\begin{eqnarray*}
H_1 : \mathrm{\log_2} \frac{VIc(x)}{VIc_{random}(x)} \sim \mathcal{GP}(m(x), k(x,x\textprime)).
\end{eqnarray*}

The marginal likelihood derived from Eq. (\ref{GPmarginal}), enables then to compare or rank different models by calculating the Bayes Factor (BF). 
More specifically the BF is approximated with a log-ratio of marginal likelihoods of two GPs, each one representing the hypothesis of differential (the profile has a significant underlying signal) and non differential expression (there is no underlying signal in the profile, just random noise). 
The significance of the profiles is then assessed based on the BF. 

\subsection{Functional Principal Component testing}\label{sec:FPC}
In this section we briefly summarise the approach proposed in \cite{Pomann2016} to test the hypothesis (\ref{H0_fadtest}) when the observed data are realisations of the curves at finite grids and possibly corrupted by noise. Their motivating application is a diffusion tensor imaging study, where the objective is to compare white matter track profiles between healthy individuals and multiple sclerosis patients. They introduce a novel framework based on functional principal component analysis ($FPCA$) of an appropriate mixture process, referred to as marginal $FPCA$. The statistical framework for this problem assumes to observe data arising from two groups, namely  $\{ ( t_{1ij}, Y_{1ij}): j=1\ldots m_{1i}\} _{i=1}^{n_1} $ and  $\{ ( t_{2ij}, Y_{2ij}): j=1\ldots m_{2i}\} _{i=1}^{n_2} $, where $t_{1ij},t_{2ij} \in T$, a compact interval that in our case is T =[0,1] (time plays the role of perturbation level $p$). It is assumed that the $Y_{1ij}$  and the $Y_{2ij}$  are independent realisations of two underlying processes observed with noise on a finite grid of points:
\begin{subequations}
\begin{align*}
Y_{1ij} & =X_{1ij}+\epsilon_{1ij}  \\
Y_{2ij} & =X_{2ij}+\epsilon_{2ij} 
\end{align*}
\end{subequations}
where $X_{1ij} \sim^{IID}X_1(\cdot)$ and $X_{2ij} \sim^{IID}X_2(\cdot)$ are independent and square integrable random functions over $T$, for some underlying (latent) random processes $X_1$ and $X_2$. It is assumed that $X_1$ and $X_2$ are second-order stochastic processes with mean functions assumed to be continuous and covariance functions assumed to be continuous and positive semidefinite, both being unknown. The measurement errors $\{\epsilon_{1ij}\} $ and $\{\epsilon_{2ij}\} $ are independent and identically distributed ($IID$), with zero mean and variances $\sigma_1^2$ and $\sigma_2^2$, respectively. The authors exploit the truncated Karhunen-Lo$\grave{e}$ve expansion of the mixture process $X(\cdot)$ of $X_1(\cdot)$ and $X_2(\cdot)$
with mixture probabilities $p$ and $1-p$. Let $Z$ a binary random variable taking values in $\left\{1,2\right\}$ with $P(Z=1)=p$, then $X_1(\cdot)=E\left[(\cdot)/Z=1\right]$ and $X_2(\cdot)=E\left[(\cdot)/Z=2\right]$. Let us consider the truncated Karhunen-Lo$\grave{e}$ve expansion of $X(\cdot)$ and define $X_Z^K (t)=\mu(t)+\sum_{k=1}^K\xi_{Zk}\Phi_k (t)$, $Z=1,2$, testing hypothesis (\ref{H0_fadtest}) reduce to testing if the $FPC$ scores $ \{\xi _1^k\}_{k=1}^K$ and $\{\xi_2^k\}_{k=1}^K$ have the same distribution: 
\begin{equation}\label{H0k}
H_0^K: \{ \xi _1^k \}_{k=1}^K \buildrel d \over{=}   \{  \xi _1 ^k\}_{k=1}^K  
\end{equation} 
In practice the authors consider $K$ null hypothesis given the finite truncation level and propose a multiple two-sample univariate test, the $Anderson$-$Darling$ ($AD$) statistic \cite{Petit1976},  combined with a multiple-comparison adjustment. The authors propose a Bonferroni correction, a procedure which controls the probability of erroneously rejecting even one of the true
null hypotheses, the Family Wise Error Rate (FWER). In this case hypothesis (\ref{H0k}) is rejected if
\begin{equation}
  \min_{1\leq k \leq K} p_k \leq  \alpha / K
\end{equation}
where $p_k$ is the p-value that is obtained by using the chosen univariate two-sample test for each ${H^k_0}$.
 
The false discovery rate (FDR), suggested in \cite{BH1995} is a different point of view for how the errors in multiple testing could be considered. The FDR is the expected proportion of erroneous rejections among all rejections. If all tested hypotheses are true, controlling the FDR controls the traditional FWER. But when many of the tested hypotheses are rejected, indicating that many hypotheses are not true, the error from a single erroneous rejection is not always as crucial for drawing conclusions from the family tested, and the proportion of errors is controlled instead. Using the individual testing statistics proposed in \cite{Pomann2016} we will therefore adopt this FDR approach to adjust our tests for multiplicity.

Note that this procedure is designed for a more general framework in which the two curves $VI$ and $VIc$ can be observed at different time points (i.e. $p\in[0,1]$).

\subsection{Interval-wise Functional testing}\label{sec:IFT}
In the following we will briefly review the Interval-wise Functional testing procedure ($ITP$) proposed by \cite{Pini2016}, where the authors develop a non-parametric domain-selective inferential methodology for functional data embedded in the $L^2(T)$ space (where $T$ is any limited open interval of $\mathbb{R}$) to test (\ref{H0_fadtest}). Their technique is not only able to assess the equality in distribution between functional populations, but also to point out specific differences. They propose a procedure based on the following three steps:
\begin{enumerate}
\item Basis Expansion: functional data are projected on a functional basis (i.e. Fourier or B-splines expansion); 
\item Interval-Wise Testing: statistical tests are performed on each interval of basis coefficients; 
\item Multiple Correction: for each component of the basis expansion, an adjusted p-value is computed from the p-values of the tests performed in the previous step.
\end{enumerate}
More in detail, let us assume to observe two independent samples of sizes $n_1$ and $n_2$ of independent random functions on a separable Hilbert space $y_{ij}(t)$, $i=1,\dots,n_j$, $j=1,2$.  

In the first step, data are projected on a finite-dimension subspace generated by a reduced basis  $y_{ij}(t)=\sum_{k=1}^pc_{ij}\Phi^{(k)}(t)$, where 
integer $p$ represents the dimension. 
It follows that each of the $n=n_1+n_2$ units can be represented by means of the corresponding p coefficients $\{c_{ij}^{(k)}\}$, $k=1,...,p$; moreover, for each $k$, $c_{i1}^{(k)}$, $i=1,\dots,n_1$ and $c_{i2}^{(k)}$, $i=1,\dots,n_2$ are independent, and $c_{i1}^{(k)} \sim C_1^{(k)}$ and $c_{i2}^{(k)} \sim C_2^{(k)}$ where $C_1^{(k)}$ and $C_2^{(k)}$ denote the (unknown) distributions of the $k$th basis coefficient in the two populations.  

In the second step, the authors build a family of multivariate tests for 
$$
H_0^{(\mathbf{k})}=\cap_{k \in \mathbf{k}}H_0^{(k)}, \quad H_0^{(k)}: C_1^{(k)}\overset{d}{=}C_2^{(k)},
$$
$k=1,\dots,p$ and $\mathbf{k}$ is a vector of successive indexes in $\{1,\dots,p\}$. In addition they add the multivariate tests on the complementary sets of each interval, i.e., they do also test each hypothesis $H_0^{(\mathbf{k^c})}=\cap_{k \notin \mathbf{k}}H_0^{(k)}$. 
The tests are performed by the Nonparametric Combination Procedure (NPC), see \cite{Pesarin2010}, that constructs multivariate permutation tests by means of combining univariate-synchronized permutation tests.

In the third step they obtain the adjusted p-value for the $k$th component $\lambda_{\mathrm{ITP}}^{(k)}$
by computing the maximum over all p-values of interval-wise tests whose null hypothesis implies $H_0^{(k)}$:
$$
\lambda_{\mathrm{ITP}}^{(k)}=\max\left(\max_{\mathbf{k}\ s.t. \ k \in \mathbf{k}}\lambda^{\mathbf{(k)}},\max_{\mathbf{k^c}\ s.t. \ k \in \mathbf{k^c}}\lambda^{\mathbf{(k^c)}}\right),
$$
and prove that, if we reject the $k$th adjusted p-value $\lambda_{\mathrm{ITP}}^{(k)} \leq \alpha$, then, for any interval $\mathbf{k}$ s.t. $H_0^{(k)}$ is true $\forall k \in \mathbf{k}$, the probability of rejecting any $H_0^{(k)}$ is lower or equal to $\alpha$. This property reads interval-wise control of the FWER.

\subsection{Perturbation strategy}\label{permute}
Mimicking the approach proposed by \cite{Karrer2008} and \cite{Cutillo2012}, we restrict our perturbed networks to having the same numbers of vertices and edges as the original unperturbed network, hence only the positions of the edges change. In other words we apply a Degree Preserving Randomization. Our perturbation strategy relies on the $rewire$ function belonging to the $R$ package $igraph$, using the option $keeping\_degseq$.

Moreover, we expect that a network perturbed by only a small amount has just a few edges moved in
different communities, while a maximally perturbed network produces completely random clusters. In \cite{Karrer2008} the perturbation strategy is achieved by removing each edge with a certain probability $\alpha$ and replacing it with another edge between a pair of vertex $(i,j)$ chosen at random with a probability proportional to the degree of $i$ and $j$. Our perturbation strategy consists in randomly rewiring a percentage $p$ of edges while preserving the original graph's degree distribution. The rewiring algorithm indeed chooses two arbitrary edges in each step 
(e.g. $(a,b)$ and $(c,d)$) and substitutes them with $(a,d)$ and $(c,b)$, if they do not already exists in the graph. The algorithm does not create multiple edges.  

A null percentage of permutation $p=0$ corresponds to the original unperturbed graph,
while $p=1$ corresponds to the maximal perturbation level. Varying the percentage $p$ from $0$ (original graph) to $1$ (maximal perturbation), many perturbed graph are generated and compared to the partition on the original graph by means of $VI$. Indeed we generated $10$ perturbed graph for each different level of $p \in [0,1]$. Then, from each of the obtained graphs, we generated other $10$ graphs rewiring $1\%$ of edges each time. Hence resulting in 100 graphs for each level of $p \in [0,1]$. In our setting we chose $20$ levels of $p$.

\section{Results}
\label{Results}
\begin{figure}%
\center
\includegraphics[width=0.8\textwidth]{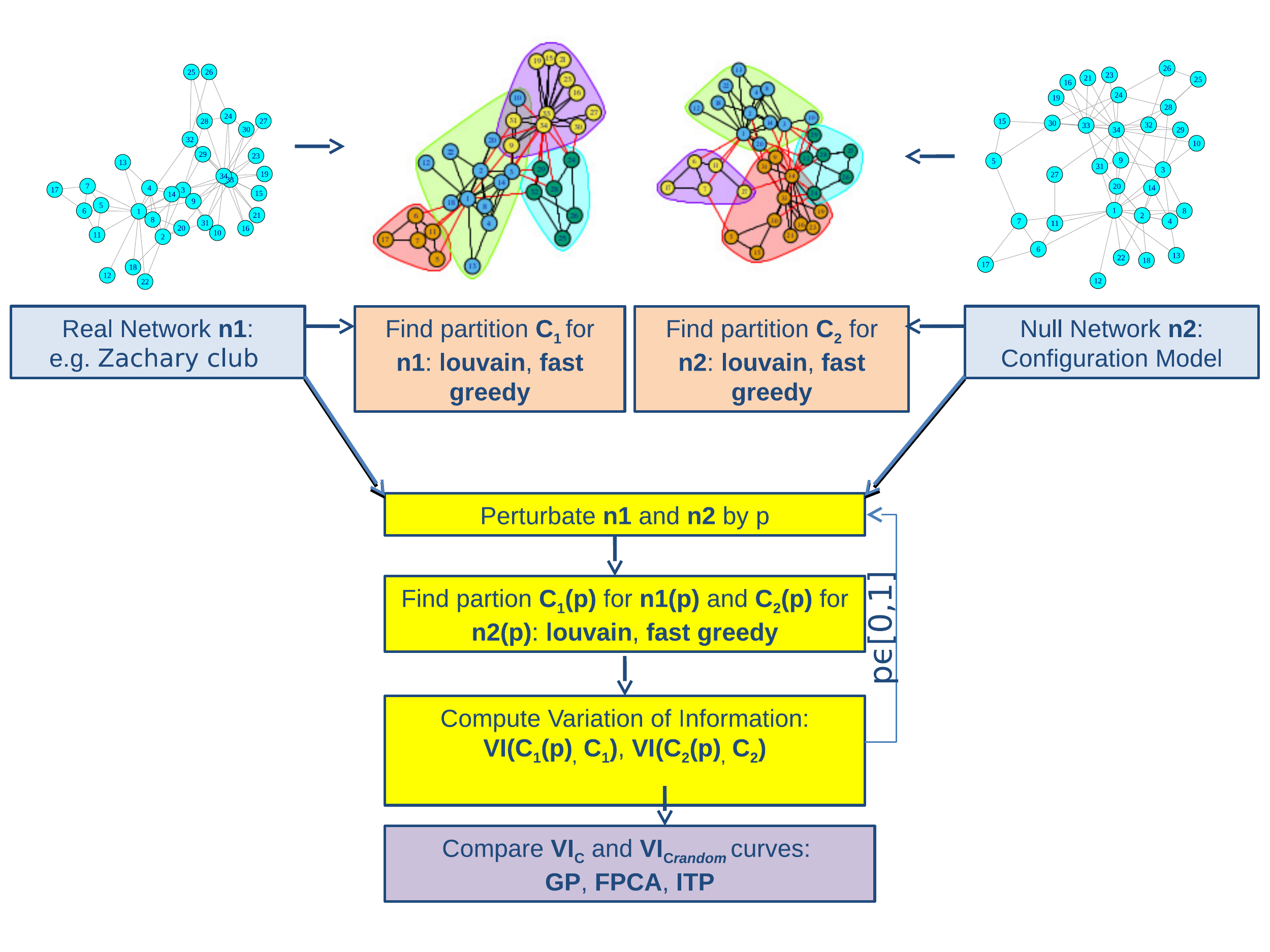}
\caption{Overall procedure map}
\label{map}
\end{figure}

The overall procedure proposed in the present paper was implemented in $R$ and validated both on simulated and real networks as will be described in the following Sections \ref{SD} and \ref{RD}. In Figure \ref{map} we provide a flow chart that summarises our procedure. For each of the analysed networks (either simulated or real) we performed the first step of the overall procedure described in Section \ref{VI}, using some tools embedded in the $R$ package $igraph$.  We chose $igraph$  because it provides an implementation of graph algorithms able to fast identifying community structures in  large graphs. In particular we used two community extraction functions, one based on a greedy optimisation of the modularity ($cluster\_fast\_greedy$) and another based on a multi-level optimisation of the modularity ($cluster\_louvain$). More specifically $cluster\_fast\_greedy$ implements the hierarchical agglomeration algorithm for detecting community structure described in \cite{Clauset2004} and  $cluster\_louvain$ is based on the hierarchical approach proposed in \cite{Blondel2008}. Both these methods enables for an automatic definition of the optimal number of communities, are specific for large networks and are based on the optimisation of the modularity. They are briefly summarised in the following.
\newline

As regards the testing methodologies we used the bioconductor package gprege available at\\
\url{https://www.bioconductor.org/packages/release/bioc/html/gprege.html} for the GP regression, the R code from the professor Staicu's web-site\\
 \url{http://www4.stat.ncsu.edu/~staicu/} for the Functional Principal Component test and the R package fdatest available at \\
  \url{https://cran.r-project.org/web/packages/fdatest/index.html} for the Interval-wise Functional test. 

\subsection*{Fast greedy method}
$Fast$ $greedy$ is the modularity optimization algorithm introduced by Clauset, Newman and Moore \cite{Clauset2004}. This method is essentially a fast implementation of a previous technique proposed by Newman \cite{Newman2004}. Starting from a set of isolated nodes, the links of the original graph are iteratively added such to produce the largest possible increase of the modularity. Adding a first edge to the set of disconnected vertices reduces the number of groups forming a new partition of the graph. The edge is chosen such that this partition gives the maximum increase (minimum decrease) of modularity with respect to the previous configuration. All other edges are added based on the same principle. At each iteration step, the variation of
modularity given by the merger of any two communities of the running partition is computed and the best merger chosen. The fast version of Clauset, Newman and Moore, which uses more efficient data structures, has a complexity of $O(N \log_2 N)$ on sparse graphs.

\subsection*{Louvain method}
$Louvain$ method is the fast modularity optimization by Blondel et al. \cite{Blondel2008}. This technique consists of two steps, executed alternatively. Initially, each node is in its own community.  In step 1, nodes are considered one by one, and each one is placed in the neighbouring community (including its own) that maximizes the modularity gain. This is repeated until no node is moved (the obtained decomposition provides therefore a local optimization of Newman-Girvan modularity). After a partition is identified in this way, in step 2 communities are replaced by super-nodes, yielding a smaller weighted network where two super-nodes are connected if there is at least an edge between vertices of the corresponding communities. The two steps of the algorithm are then repeated until modularity (which is always computed with respect to the original graph) does not increase any further. 

As pointed out in \cite{Fortunato2010}, this method offers a fair compromise between the accuracy of the estimate of the modularity maximum, which is better than that delivered by greedy techniques like the one by Clauset et al. above, and computational complexity, which is essentially linear in the number of links of the graph.

\subsection{Application to simulated data} \label{SD}
In order to show the ability of our method to validate a network clustering, we applied it to modular random network graphs generated using the model implemented in \cite{Sah2014}. The model generates undirected, simple, connected graphs with prescribed degree sequences and a specified level of community structure, while maintaining a graph structure that is otherwise as random (uncorrelated) as possible over a broad range of distributions of network degree and community size. The model in \cite{Sah2014} is specified by the network size, the average network degree, the number of modules, the modularity, the degree distribution and the module size distribution. The generated graph results also to be as random as possible, to contain no self loops (edges connecting a node to itself), multi-edges (multiple edges between a pair of nodes), isolate nodes (nodes with no edges), or disconnected components (see \cite{Sah2014} for details). Specifically, we generated a modular random graph for each level of modularity Q = 0, 0.2, 0.4, 0.6, 0.8, using a power law for degree distribution and for module size distribution, with size=2000,  number of modules=10 and average degree=10. For each graph, the corresponding null model was generated using the Configuration Model. \\

The application of the overall procedure on the simulated datasets is summarised in Tables \ref{BFsimul} and \ref{FADsimu} and in Figures \ref{fig:VIfastSimu} and \ref{fig:VIlouvainSimu}, respectively. 

\subsubsection*{Gaussian Process results}
The application of the Gaussian Processes approach described in Section \ref{sec:GP} to the simulated networks is summarised in Table \ref{BFsimul}. 

\begin{table}[h]
\centering
\begin{tabular}{ | l | c | c |}
\hline
   Q value & fast greedy & Louvain \\  \hline
   0   & 3.194   & 8.356   \\ \hline
   0.2 & 3.959   & 59.589  \\ \hline
   0.4 & 257.696 & 331.810 \\ \hline
   0.6 & 389.301 & 421.305 \\ \hline
   0.8 & 443.650 & 454.558 \\ \hline
\end{tabular}
\caption{GP Bayes Factor on the simulated networks with modularity Q $\in \{0,0.2,0.4,0.6,0.8\}$ after clustering via fast greedy and Louvain} \label{BFsimul}
\end{table}

The resulting BF show a growing trend from modularity $Q=0$ to $Q=0.8$  after clustering with either clustering  $fast$ $greedy$ or $louvain$. 
This gives strong statistical evidence that networks with a high modularity have a robust clustering structure (significantly different from the random).
However, note that $louvain$  method is able to recover a non random clustering also at low modularity ($Q\geq 0.2$), the other way around $fast$ $greedy$ needs a more defined structure ($Q\geq 0.4$).

\subsubsection*{Functional Principal Component testing results} 
Similarly, Table \ref{FADsimu} summarises the application of the Functional $Anderson$-$Darling$ test described in Section \ref{sec:FPC} to the simulated data. 

\begin{table}[h]
\centering
\begin{tabular}{ | l | c | c |}
\hline
   Q value & fast greedy & Louvain \\  \hline
   0 & 0.1409  &  0.1184  \\ \hline
   0.2 & 0.369   &  0.00386 \\ \hline
   0.4 & 0.00016 &  0.00016 \\ \hline
   0.6 & 0.00016 &  0.00012 \\ \hline
   0.8 & 0.00016 &  8e-05 \\ \hline
\end{tabular}
\caption{FDR adjusted p-values by the FAD test procedure on the simulated networks with modularity Q $\in \{0,0.2,0.4,0.6,0.8\}$ after clustering via fast greedy and Louvain.} \label{FADsimu}
\end{table}

As we can see the False Discovery rate adjusted p-values decreases drastically when the simulated network modularity grows from $Q=0$ to $Q=0.8$, after clustering with either $fast$ $greedy$ or $louvain$ algorithms, also if $louvain$ is able to recover a non random clustering at lower modularity ($Q\geq 0.2$) than $fast$ $greedy$.
This result agree with the previous one obtained by GP.

\subsubsection*{Interval-wise Functional Testing results}
\begin{figure}[h]
\subfloat[Q=0.6]{\includegraphics[width=0.4\textwidth]{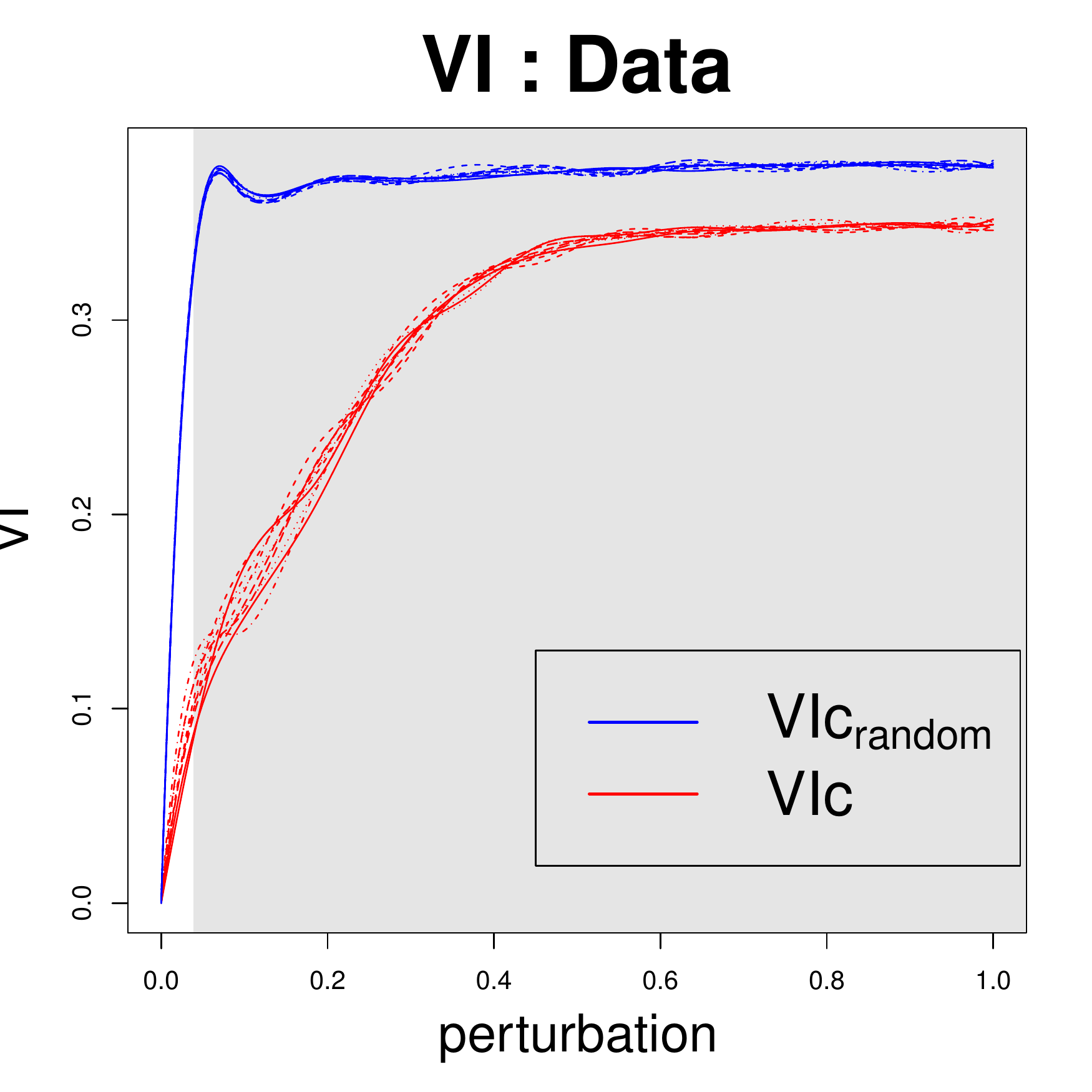}}
\subfloat[Q=0.6]{\includegraphics[width=0.4\textwidth]{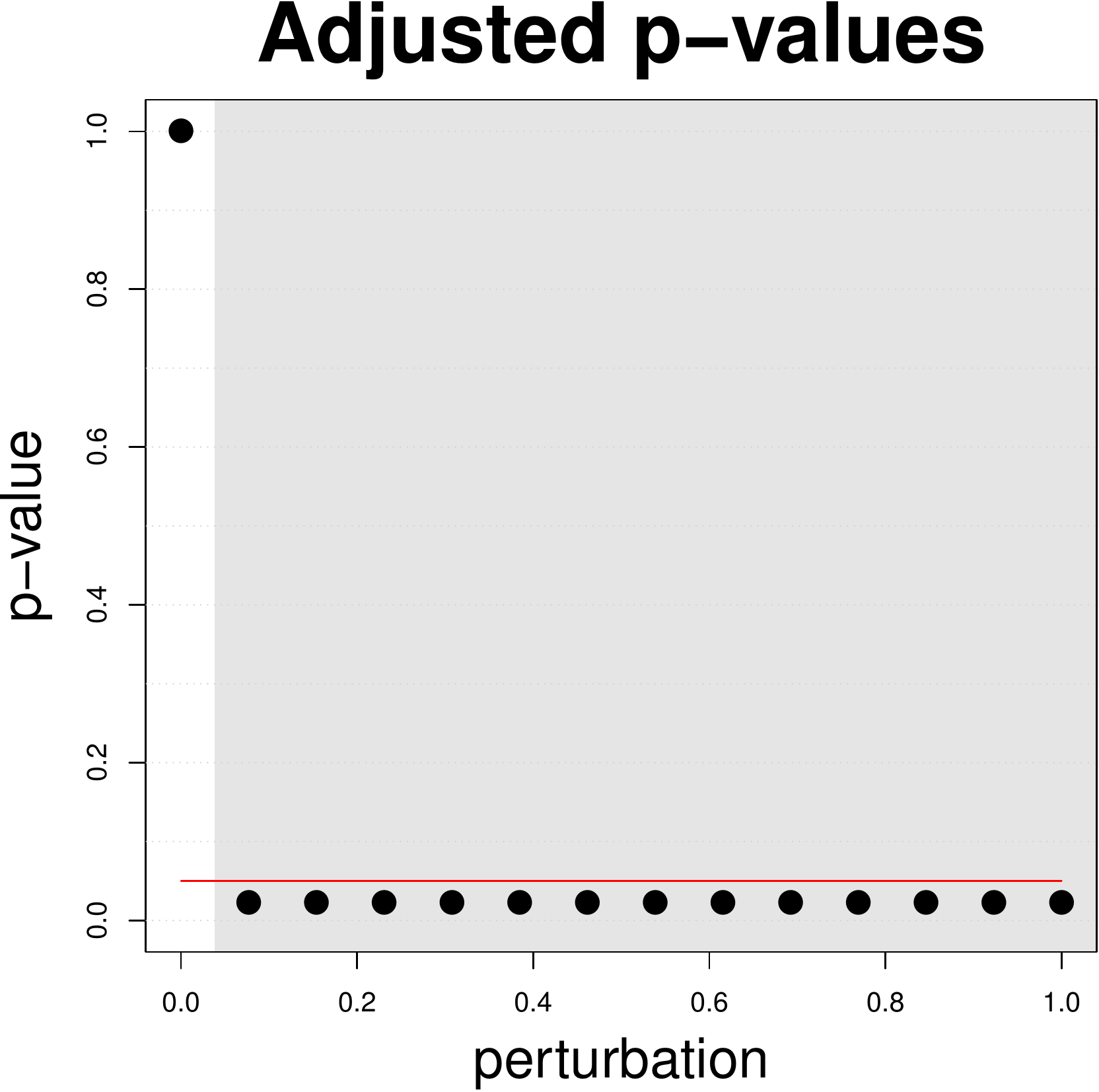}}\\
\subfloat[Q=0.8]{\includegraphics[width=0.4\textwidth]{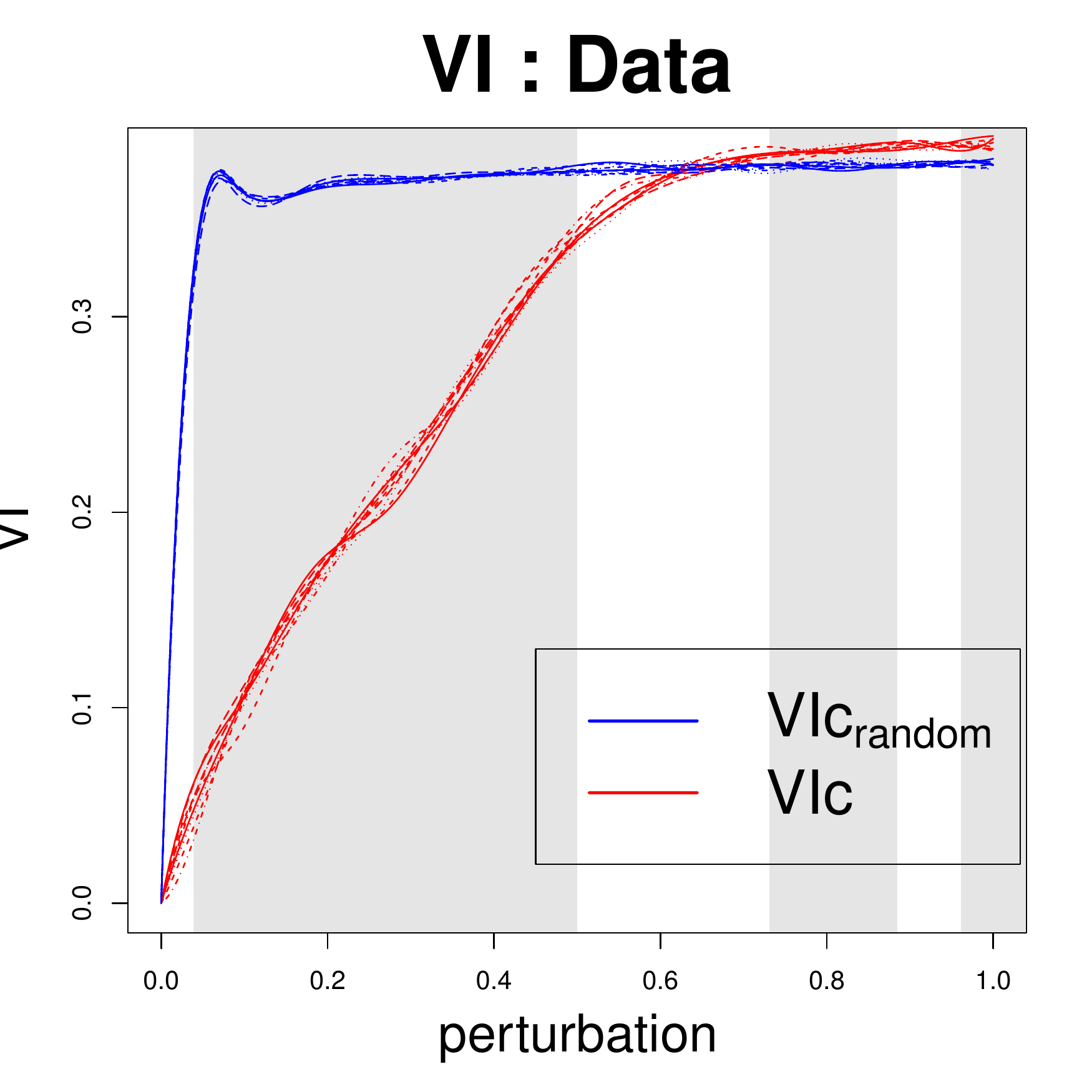}}%
\subfloat[Q=0.8]{\includegraphics[width=0.4\textwidth]{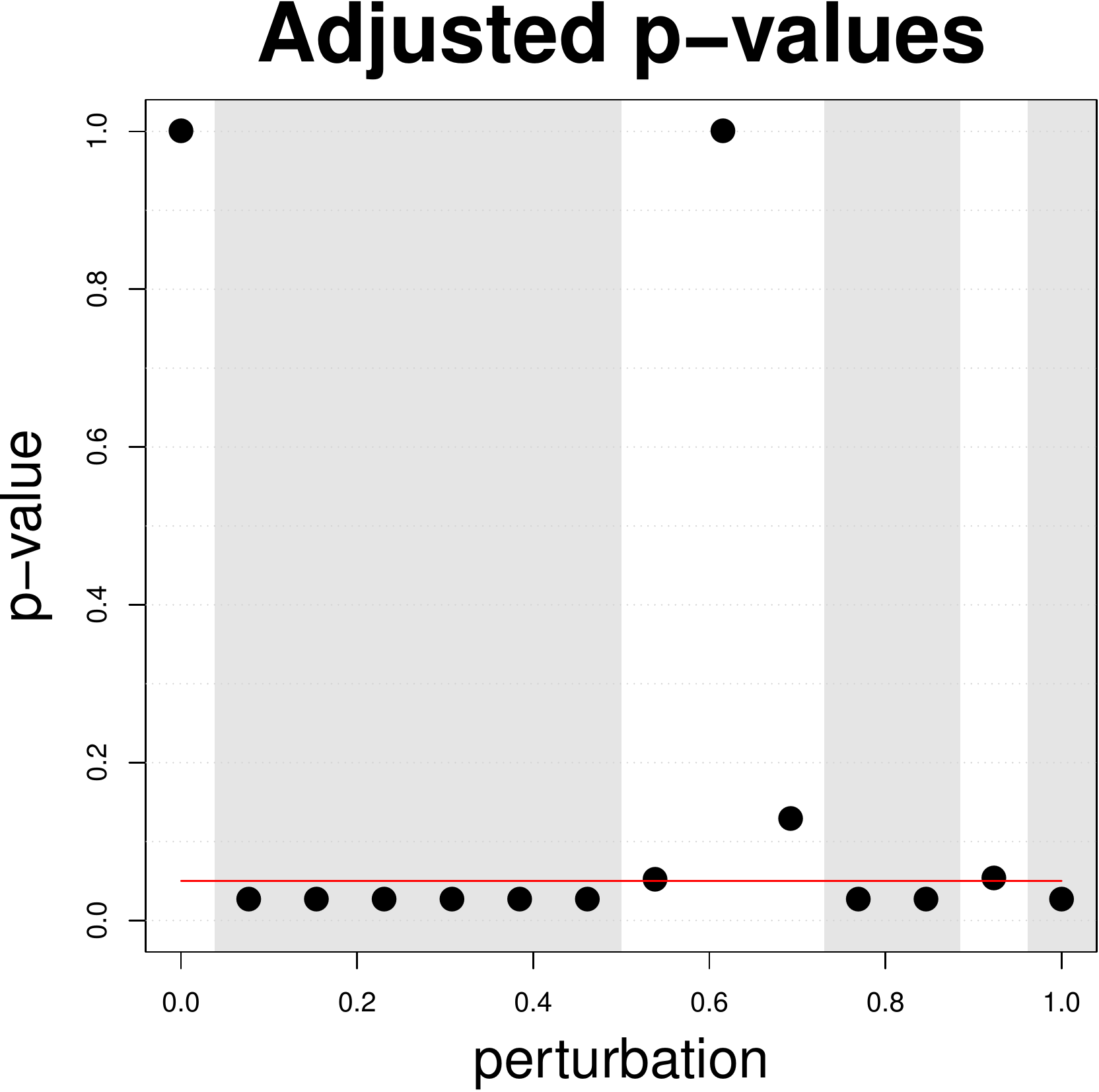}}
\caption{VI plots on the clustering obtained via fastgreedy on 5 simulated datasets with different modularity Q $\in [0,0.2,0.4,0.6,0.8]$ and corresponding adjusted p-values of the Interval Testing procedure. Horizontal red line corresponds to the critical value 0.05. Light gray areas correspond to p-values below 0.05, dark grey areas correspond to p-values below 0.01.}
\label{fig:VIfastSimu}
\end{figure}

\begin{figure}[h]
\centering
\subfloat[Q=0.6]{\includegraphics[width=0.4\textwidth]{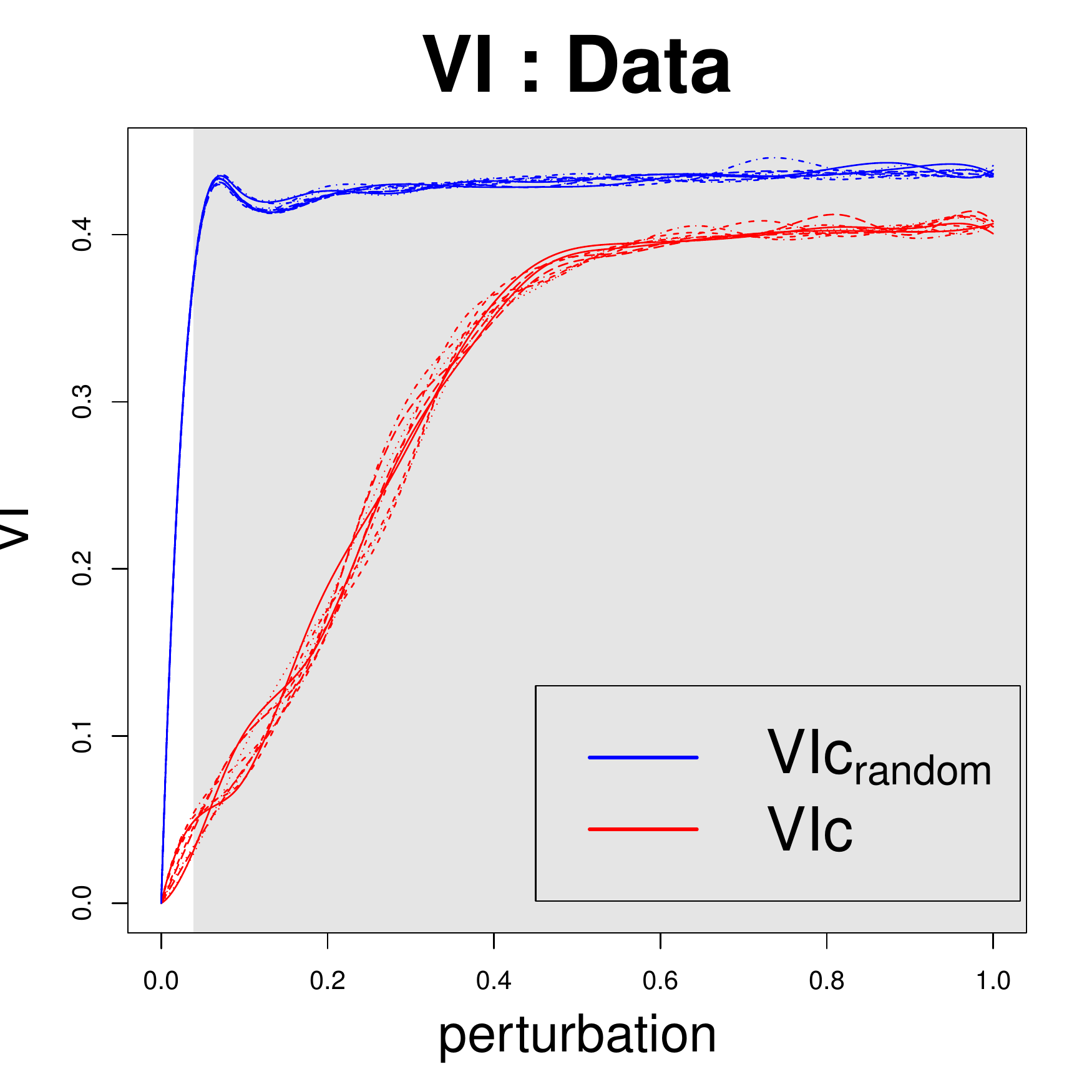}}
\subfloat[Q=0.6]{\includegraphics[width=0.4\textwidth]{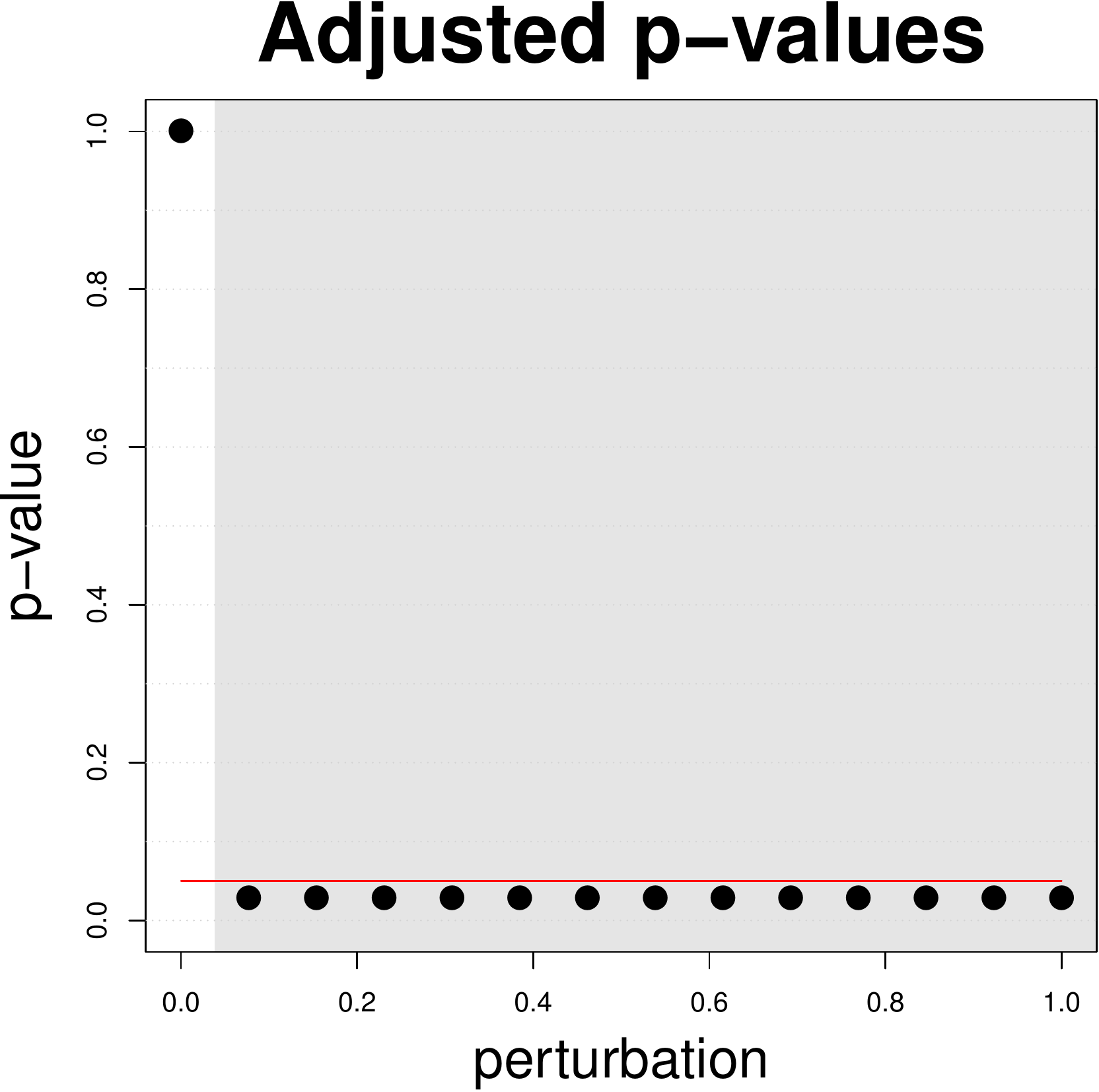}}\\

\subfloat[Q=0.8]{\includegraphics[width=0.4\textwidth]{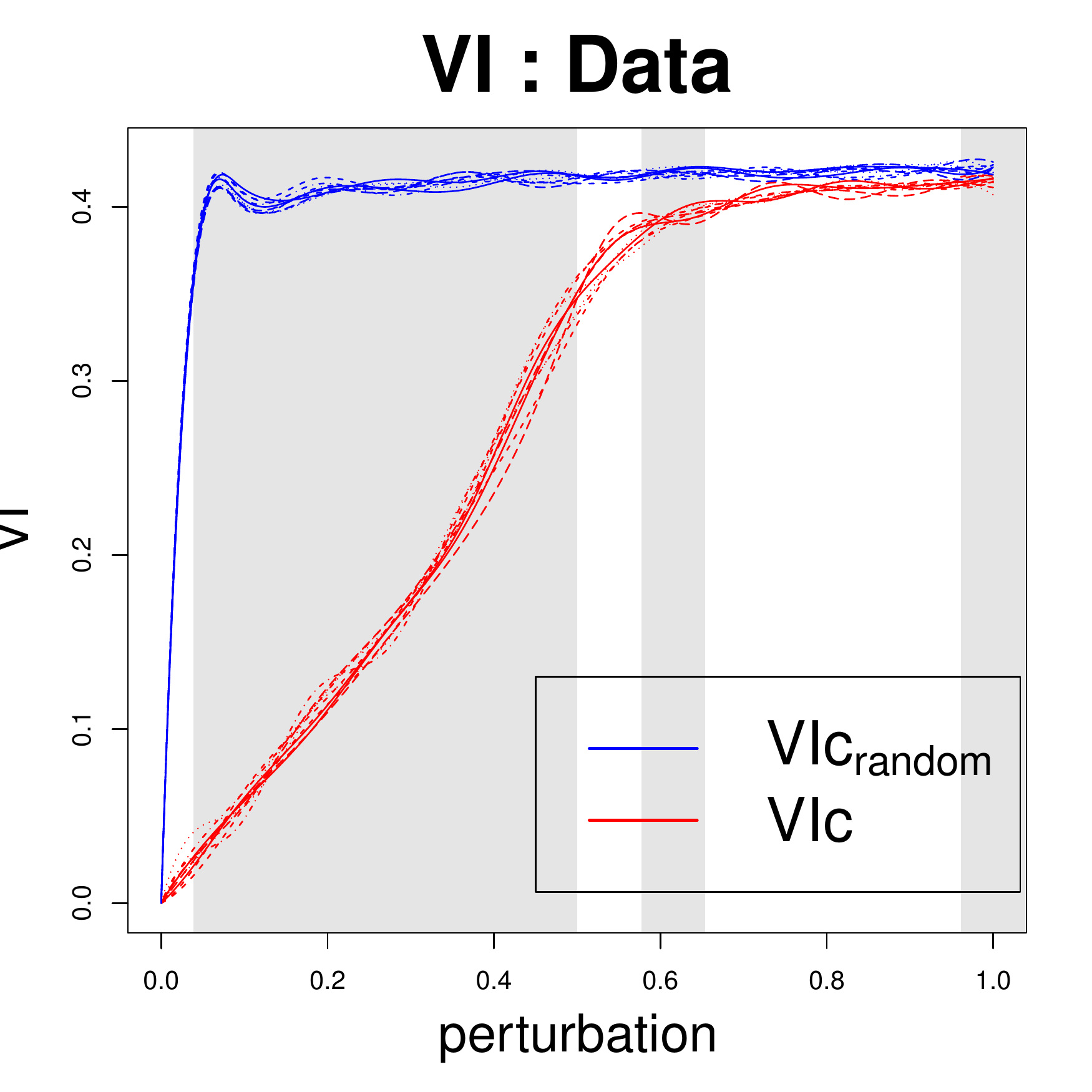}}%
\subfloat[Q=0.8]{\includegraphics[width=0.4\textwidth]{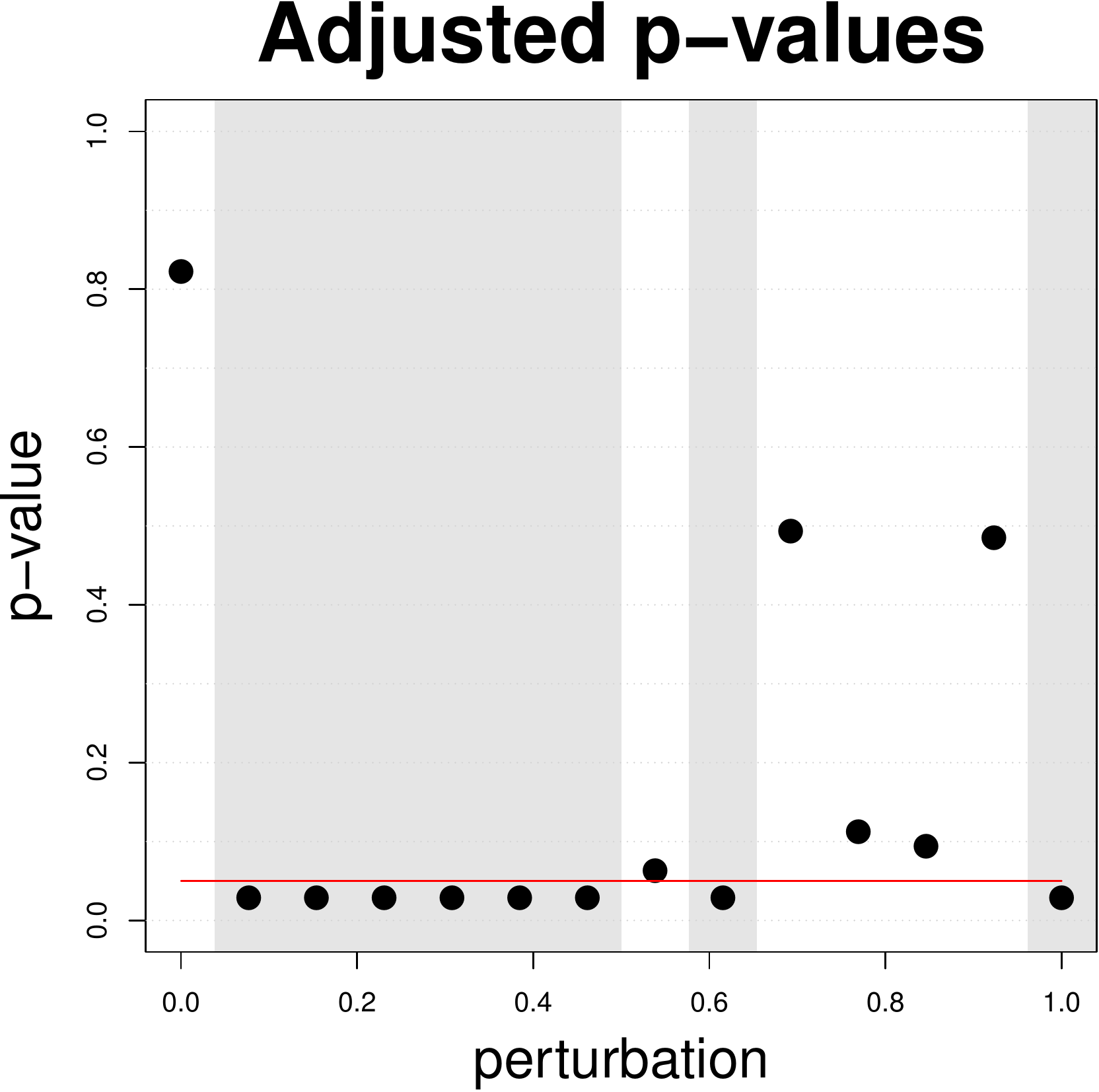}}
\caption{VI plots on the clustering  obtained via Louvain on 5 simulated datasets with different modularity Q $\in [0,0.2,0.4,0.6,0.8]$ and corresponding adjusted p-values of the Interval Testing procedure . Horizontal red line corresponds to the critical value 0.05. Light gray areas correspond to p-values below 0.05, dark grey areas correspond to p-values below 0.01.} \label{fig:VIlouvainSimu}
\end{figure}

The application of the Interval-wise Testing procedure described in Section \ref{sec:IFT} to the simulated datasets after clustering via  $fast \ greedy$ or $louvain$ are depicted respectively in Figures \ref{fig:VIfastSimu} and \ref{fig:VIlouvainSimu}.  In each figure, panels $(a),(c),(e),(g),(i)$ show the VI curves for the null model ($VIc_{random}$) and for the actual model ($VIc$). The two curves appear to be very close for low modularity values and depart from each other as the modularity increases till a maximum of $Q=1$. In panels $(b),(d),(f),(h),(j)$ this is quantified locally by a specific adjusted p-value in each sub-interval. Significant p-values are falling under the horizontal red line corresponding to the critical value of 0.05. As we can see, either using $louvain$ or $fast$ $greedy$ as clustering methods yields to the similar $ITP$ results conclusion. Of course when there is no perturbation (i.e. at level $p=0$) the two curves are coincident and hence not significantly different. When $Q\geq 0.4$ the two $VI$ curves are significantly different at any perturbation level, apart from some cases at $p\geq 0.5$ and $Q=0.8$ where the $VIc$ is close to the $VIc_{random}$, indeed note that if we strongly perturb a network ($p\geq 0.5$, i.e. we rewire more than $50\%$ of edges) it approaches a random network. Also in this case $louvain$  is able to recover a non random clustering at lower modularity ($Q\geq 0.2$) than $fast$ $greedy$, confirming the results obtained by the other two approaches. Moreover note that at $Q=0$, when applying $louvain$, the method detects some interval where there is a significant difference between the two curves but for strong perturbation level ($p\geq 0.5$).

\begin{figure}[h]%
\centering
\subfloat[Q=0]{\includegraphics[width=0.4\textwidth]{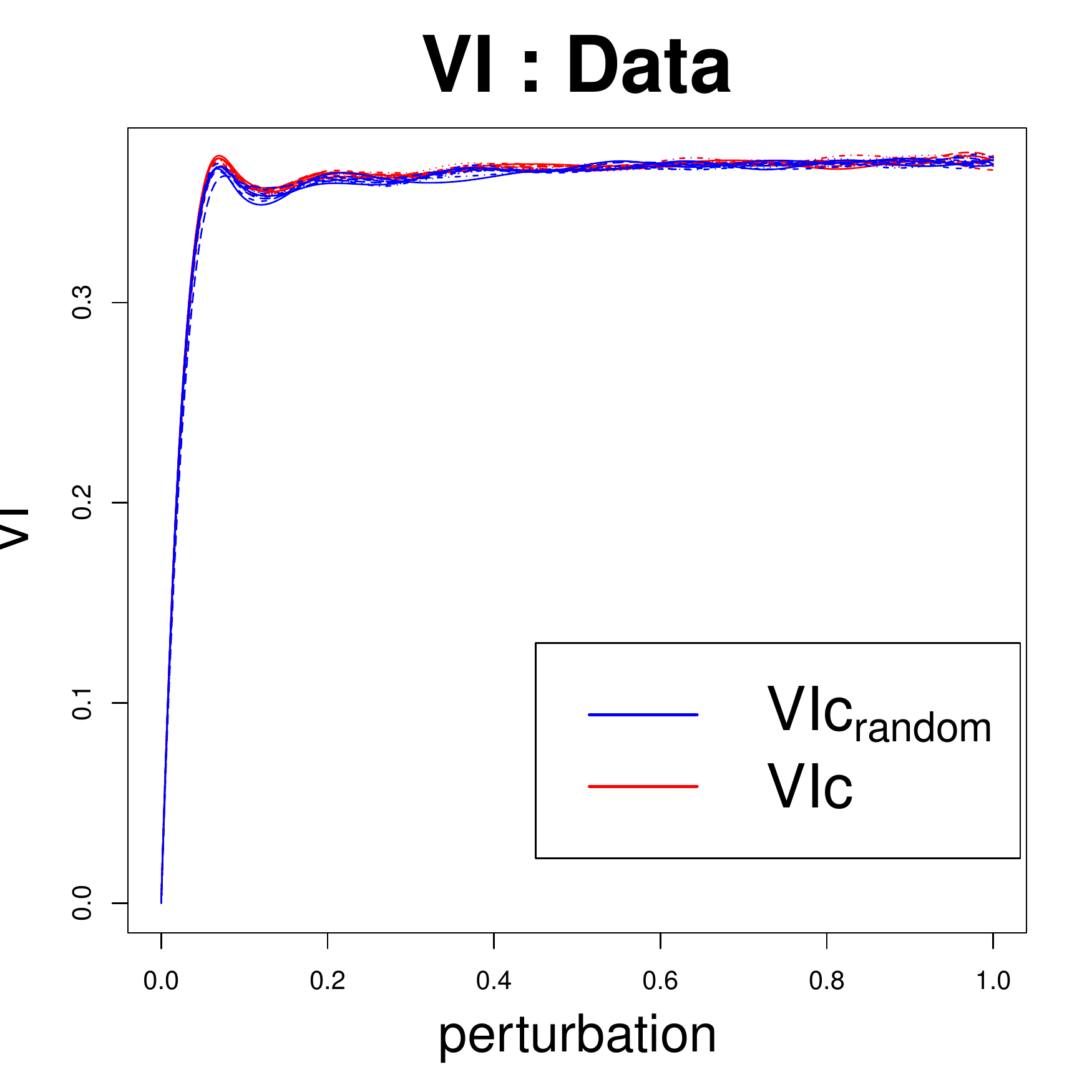}}
\subfloat[Q=0]{\includegraphics[width=0.4\textwidth]{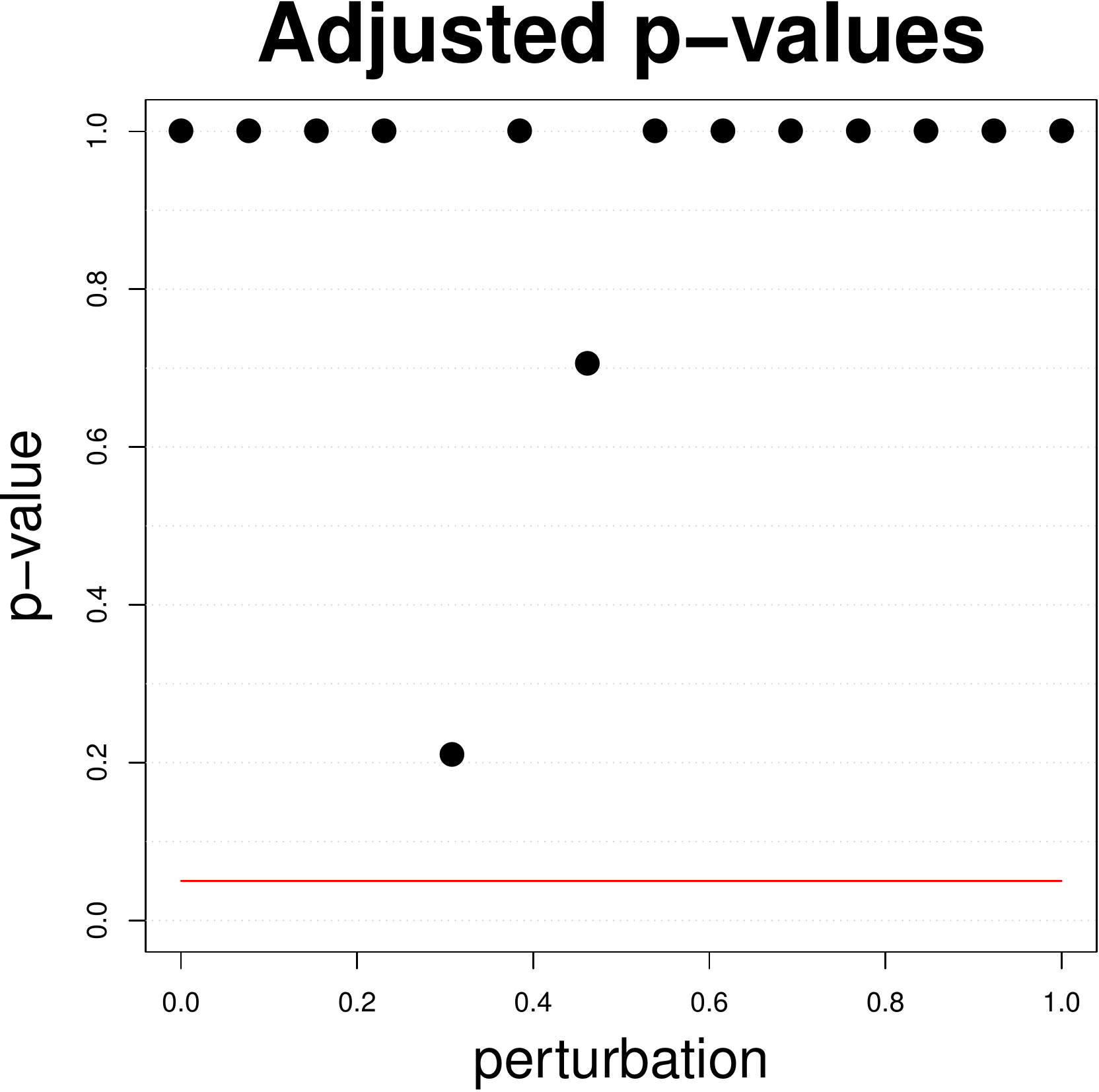}}\\

\subfloat[Q=0.2]{\includegraphics[width=0.4\textwidth]{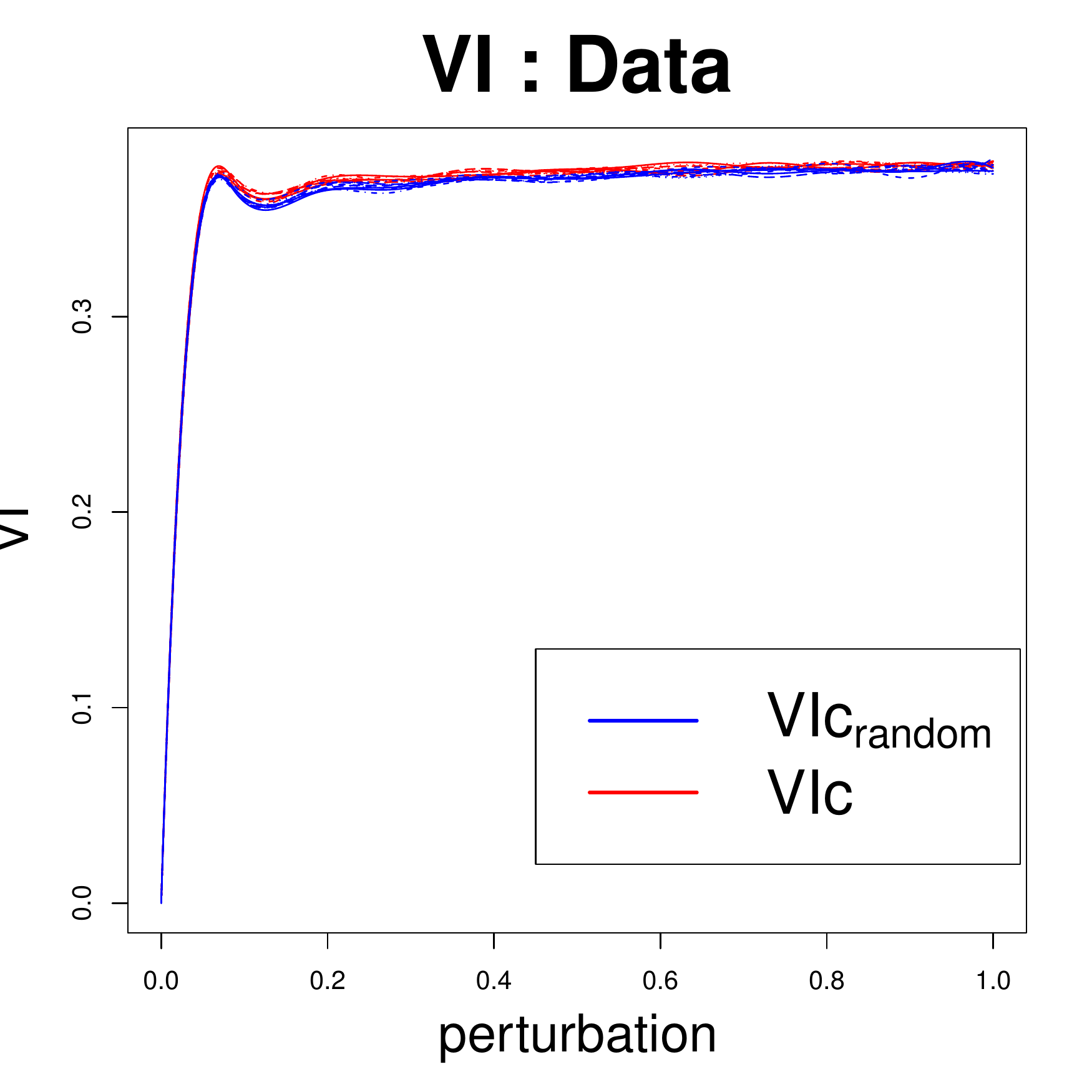}}
\subfloat[Q=0.2]{\includegraphics[width=0.4\textwidth]{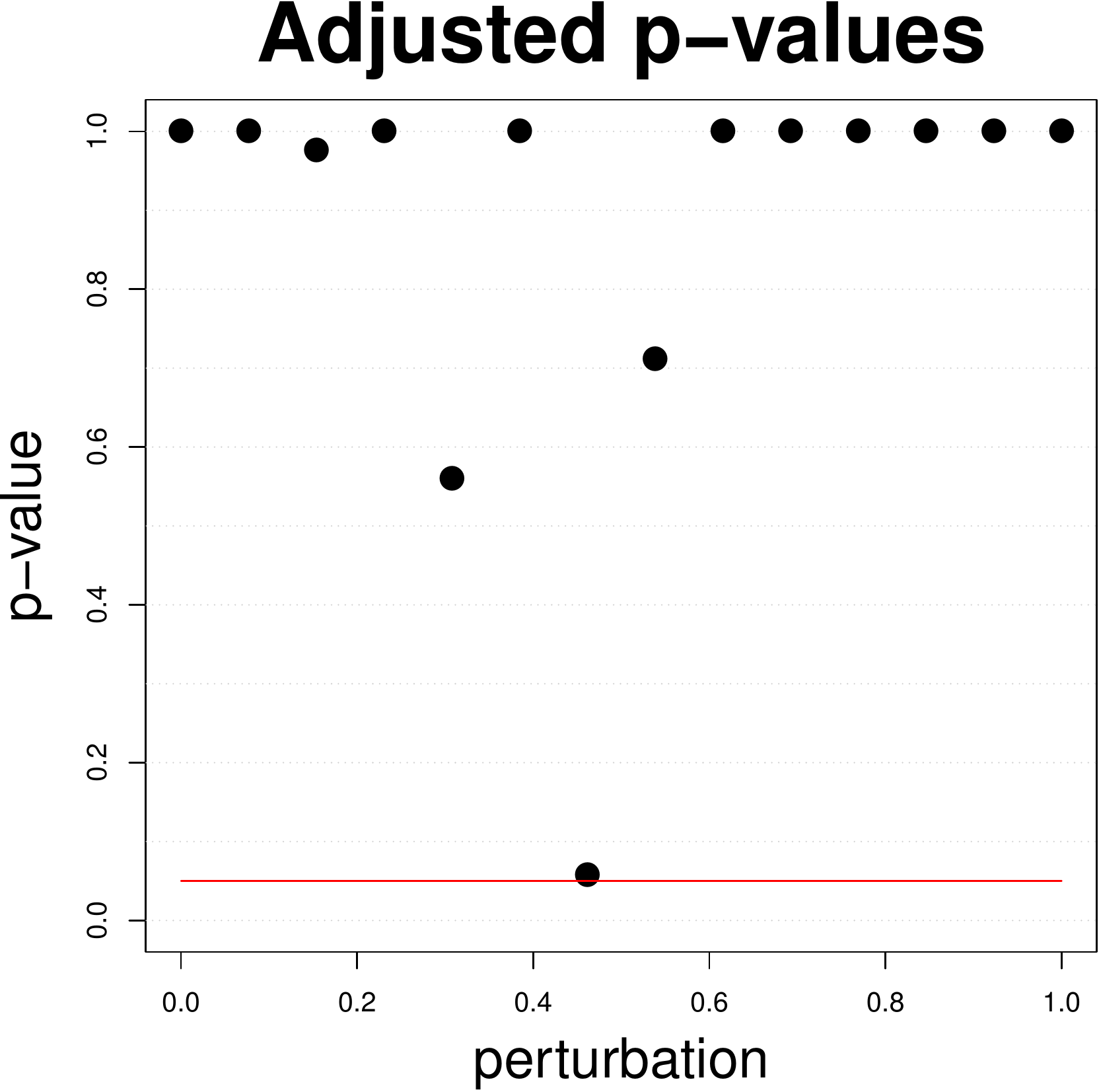}}\\

\subfloat[Q=0.4]{\includegraphics[width=0.4\textwidth]{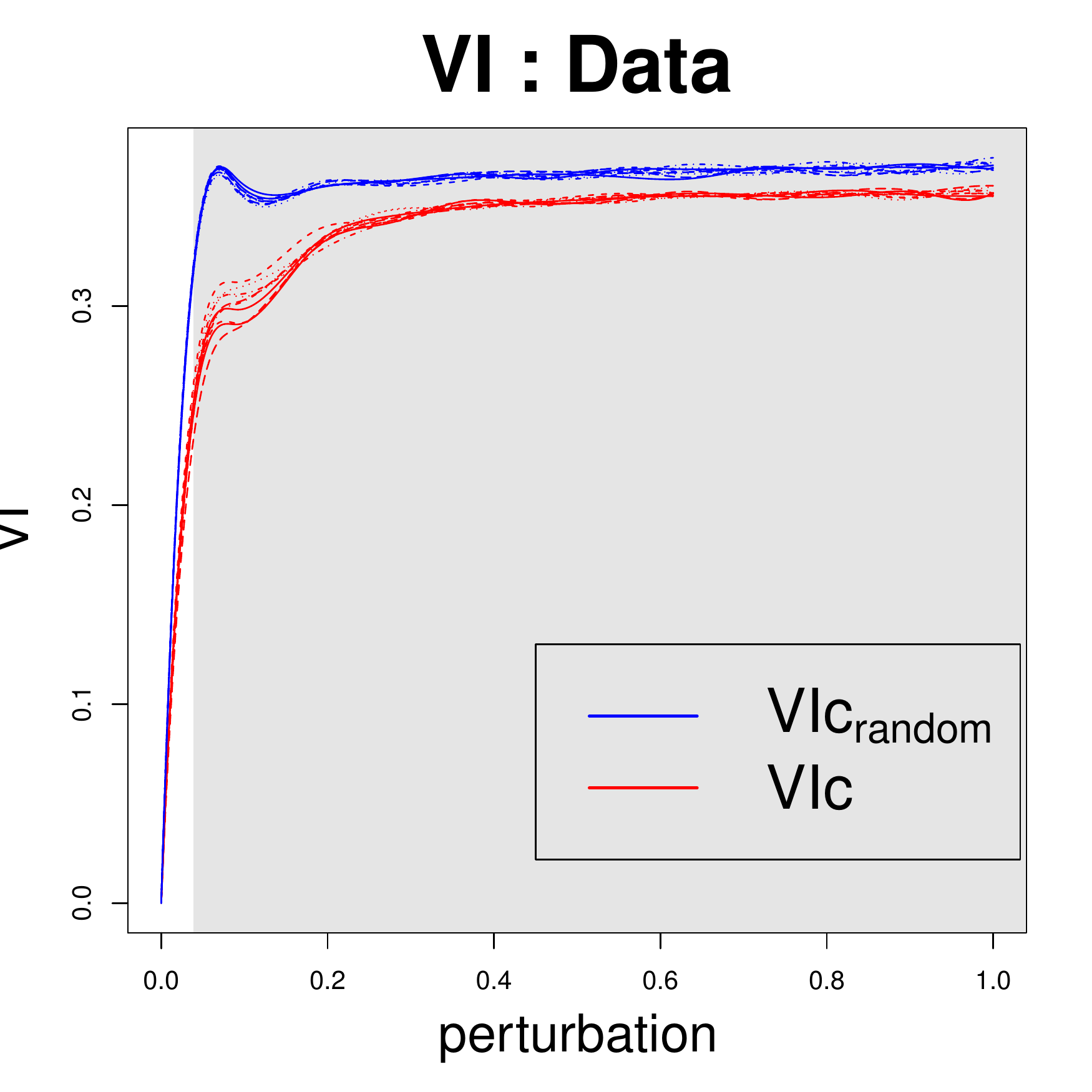}}%
\subfloat[Q=0.4]{\includegraphics[width=0.4\textwidth]{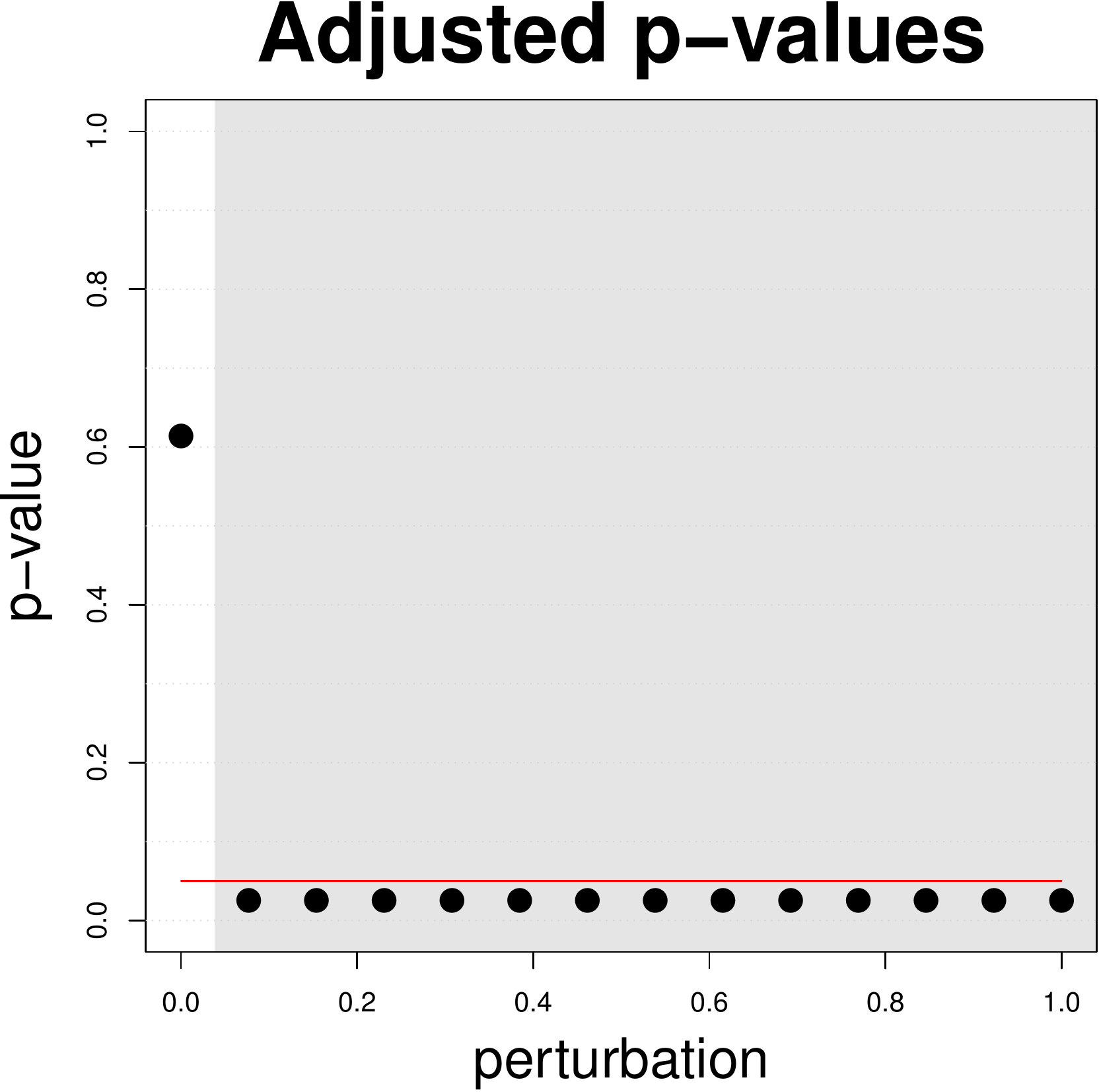}}\\
\end{figure}
\setcounter{subfigure}{6}

\begin{figure}[h]%
\centering
\subfloat[Q=0]{\includegraphics[width=0.4\textwidth]{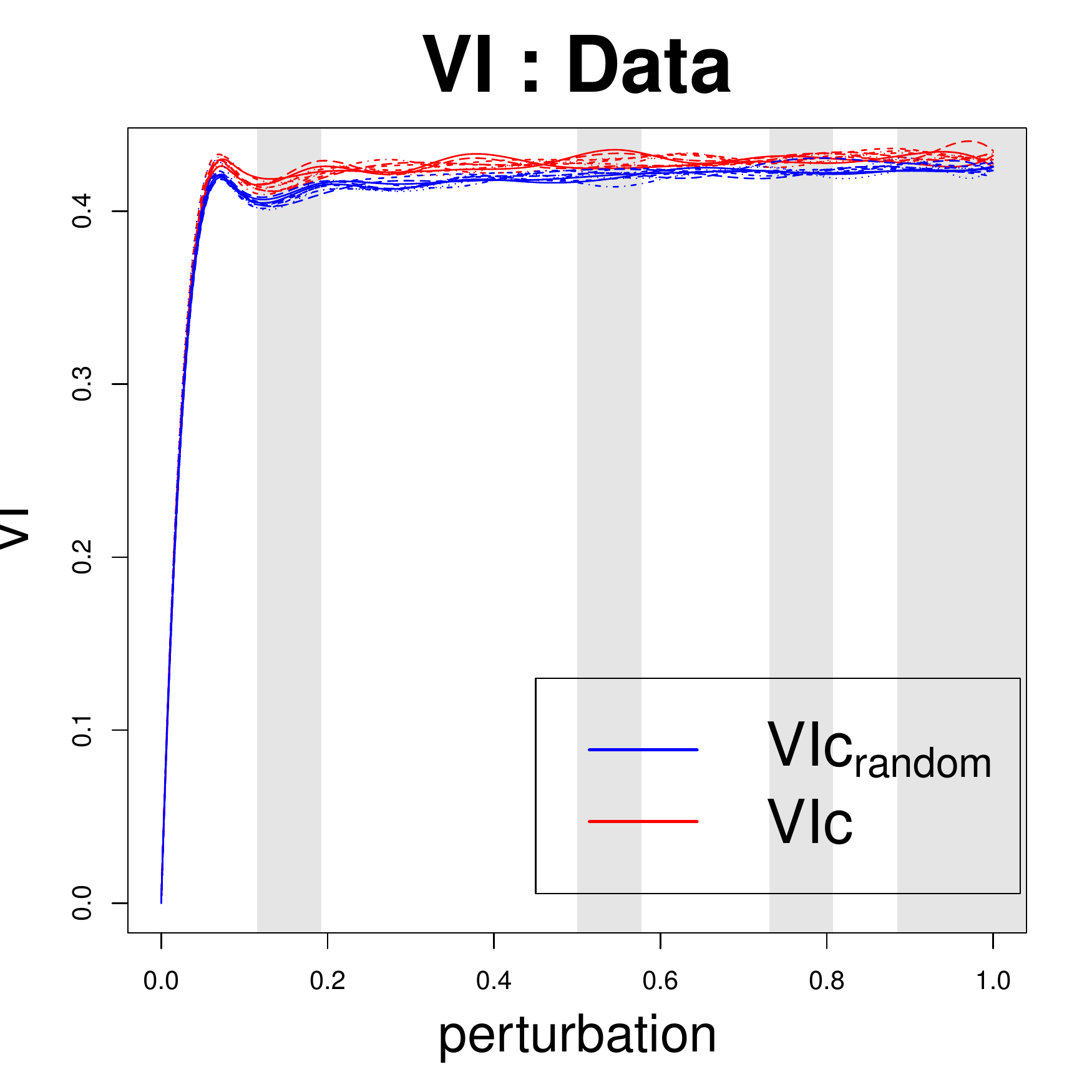}}%
\subfloat[Q=0]{\includegraphics[width=0.4\textwidth]{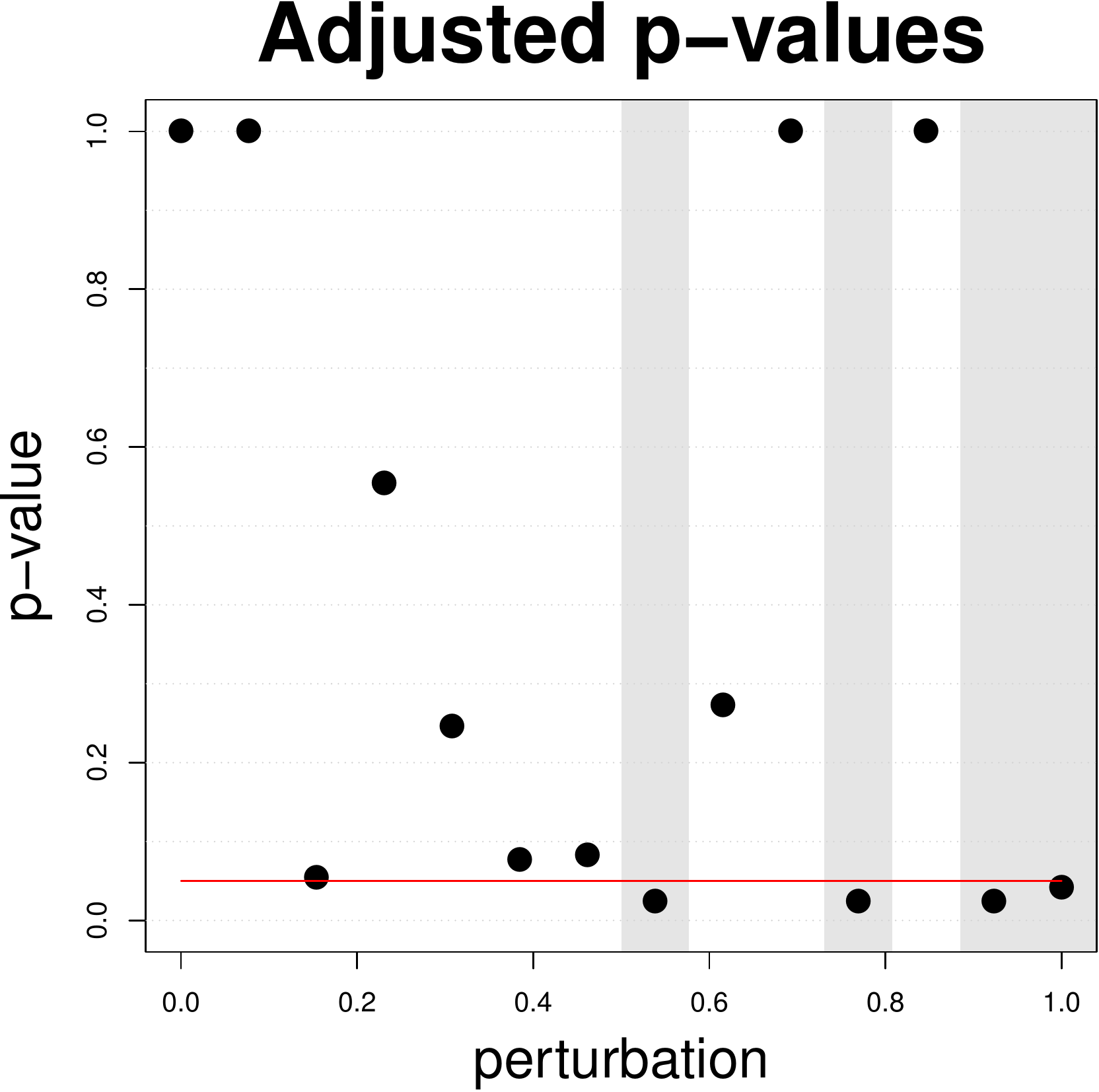}}\\

\subfloat[Q=0.2]{\includegraphics[width=0.4\textwidth]{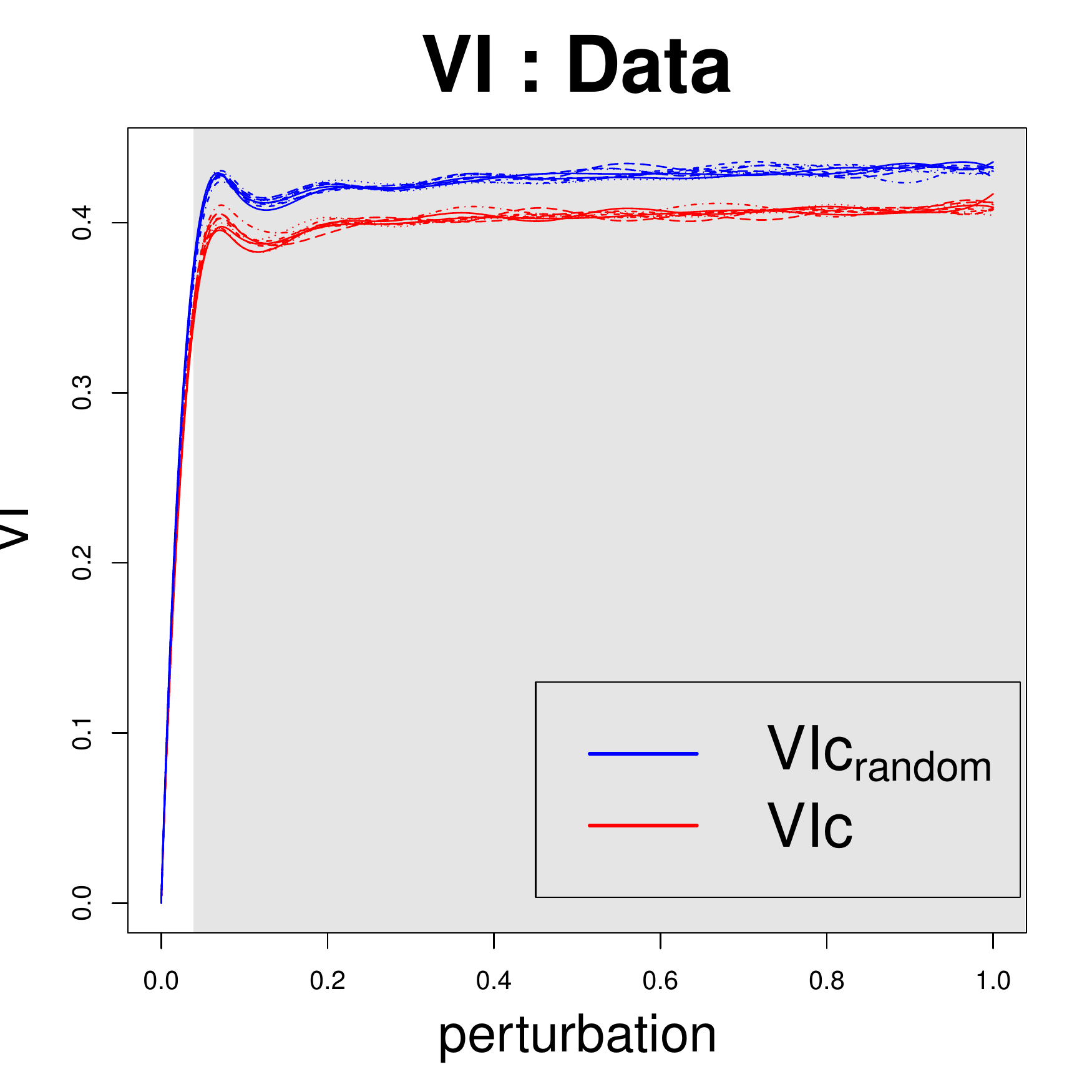}}
\subfloat[Q=0.2]{\includegraphics[width=0.4\textwidth]{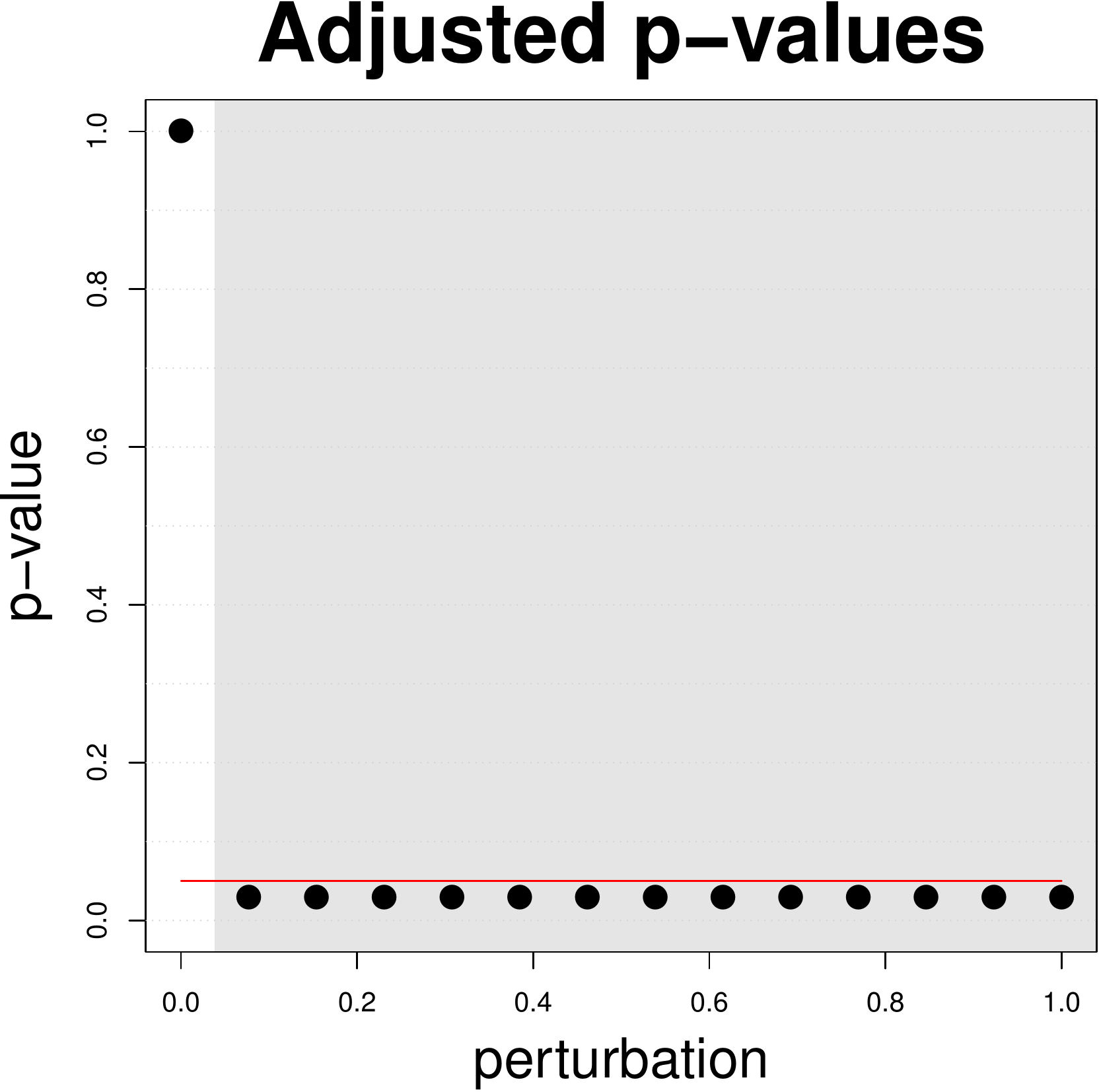}}\\
\subfloat[Q=0.4]{\includegraphics[width=0.4\textwidth]{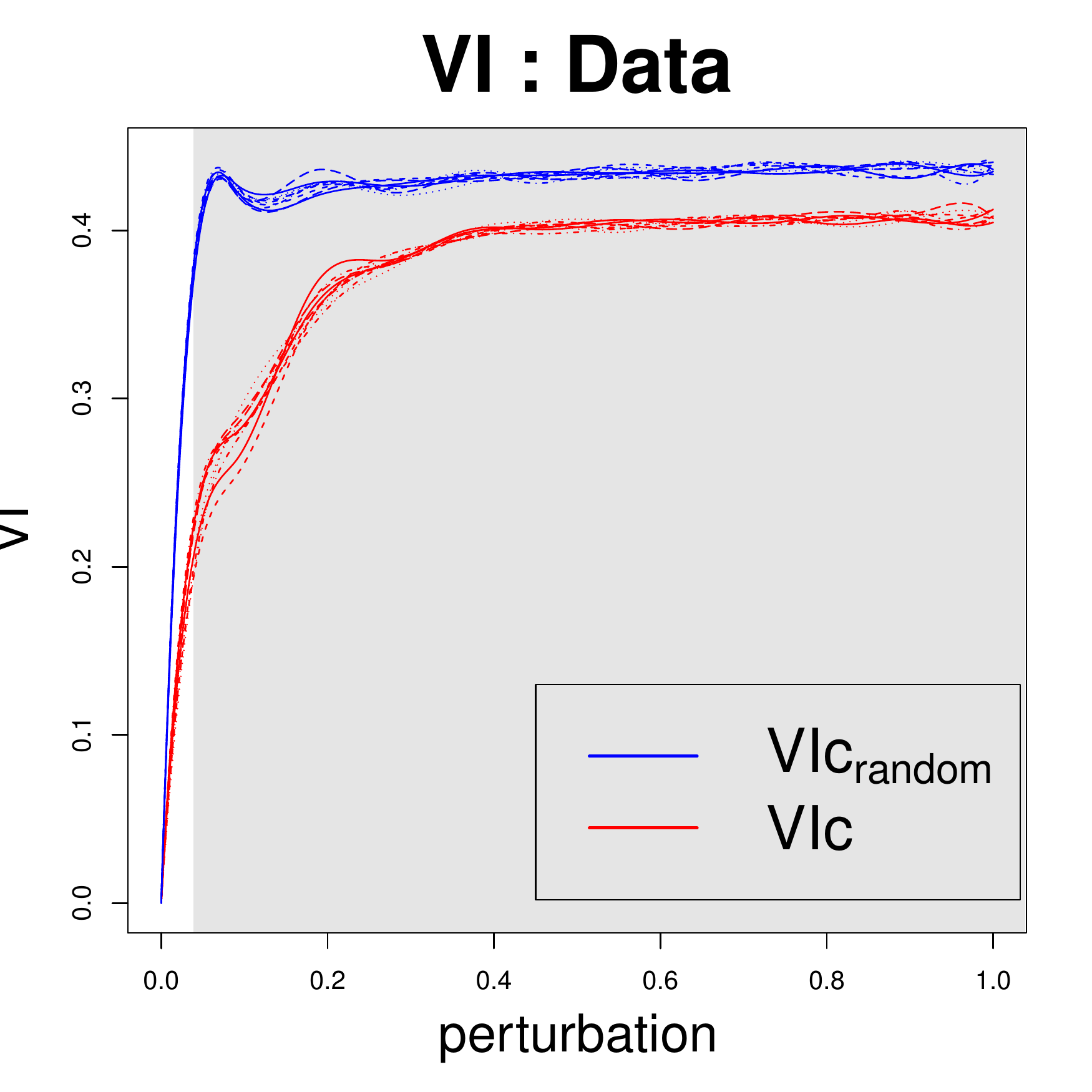}}%
\subfloat[Q=0.4]{\includegraphics[width=0.4\textwidth]{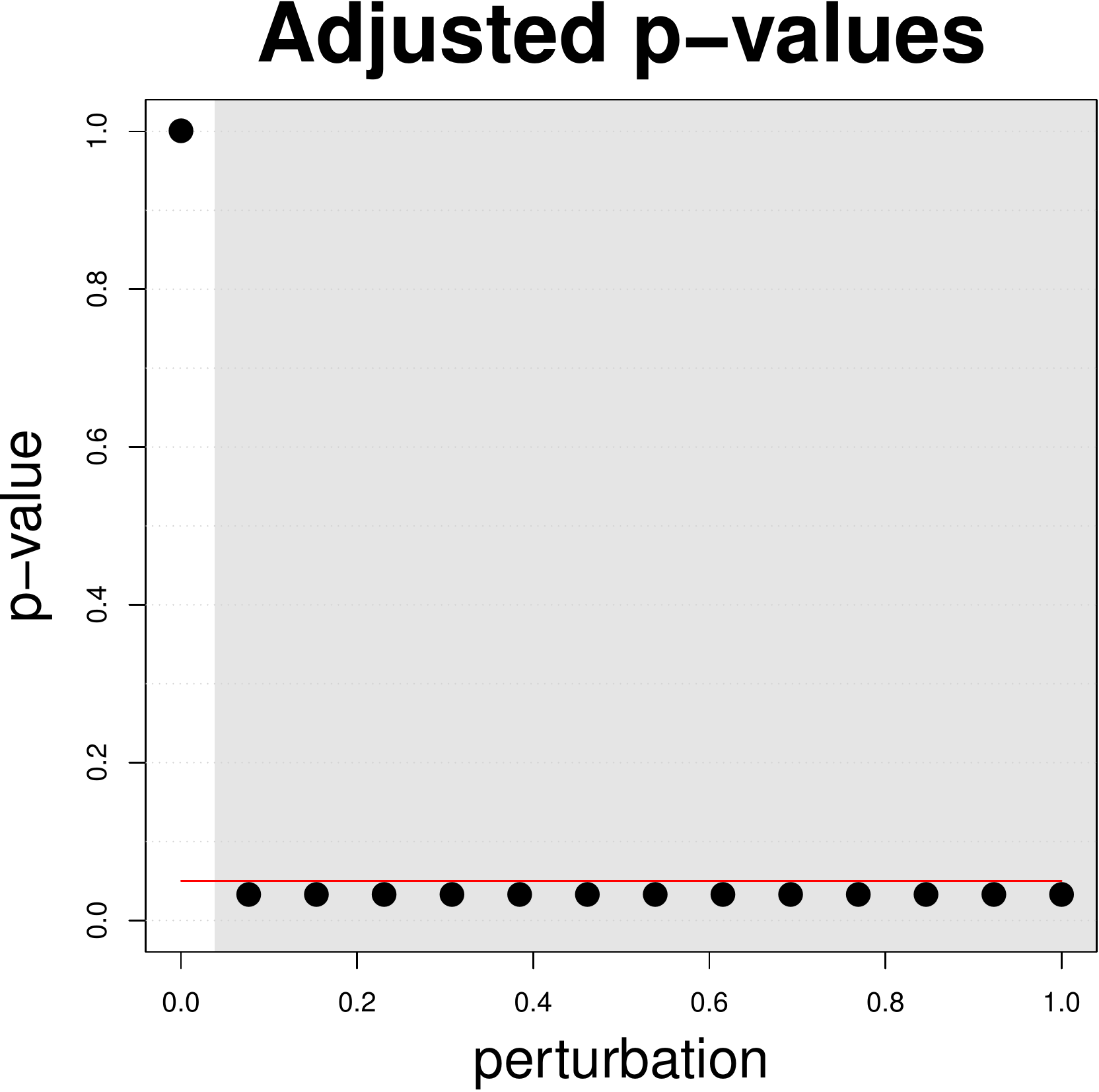}}\\
\end{figure}
\setcounter{subfigure}{6}

\subsection{Application to real data} \label{RD}
\begin{figure}[htp]
  \centering
  \caption*{Real Networks plots}

  \subfloat[Facebook]{\label{fig:facebook}\includegraphics[scale=0.4]{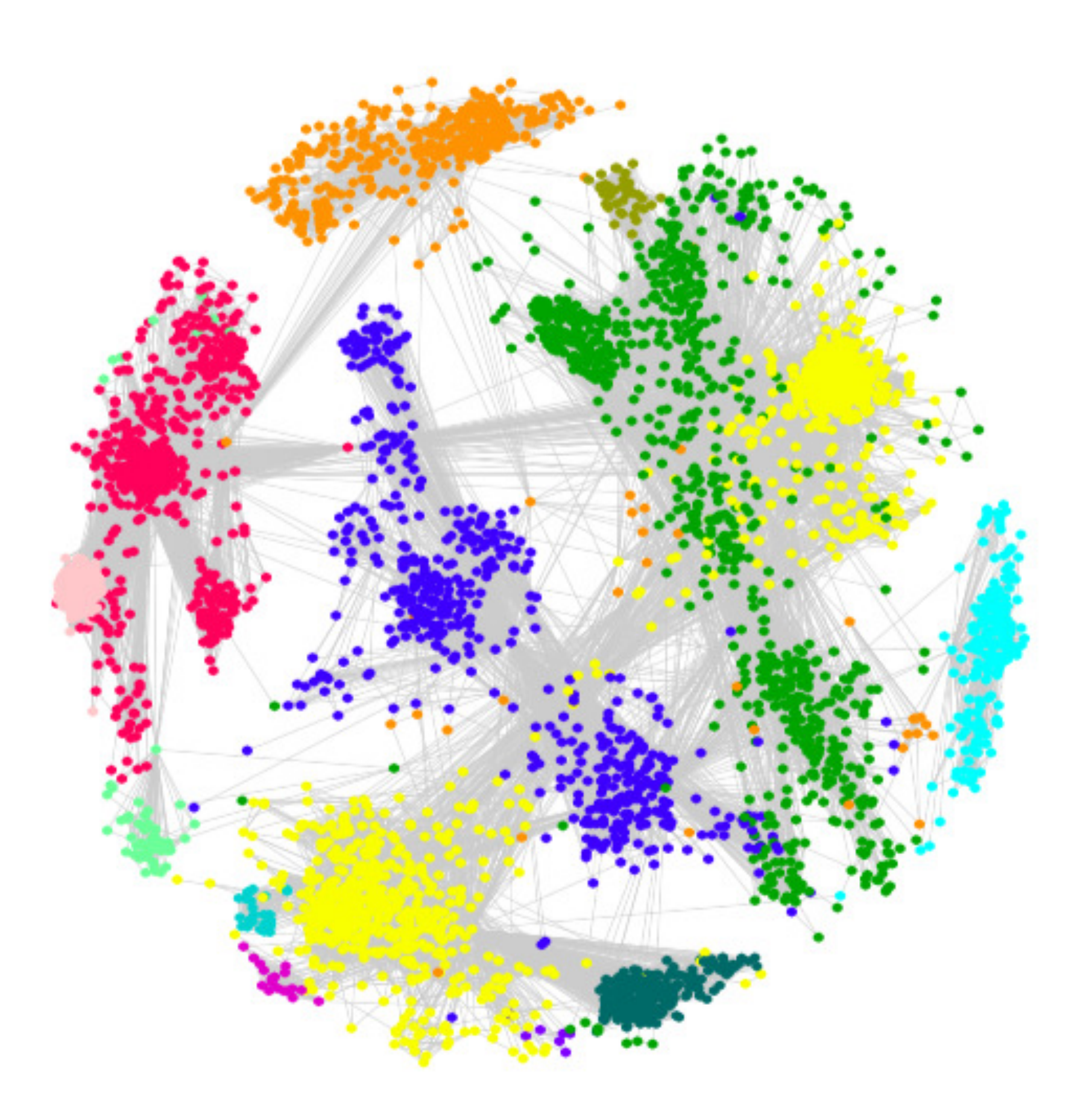}}
  \subfloat[Barabasi]{\label{fig:barabasi}\includegraphics[scale=0.4]{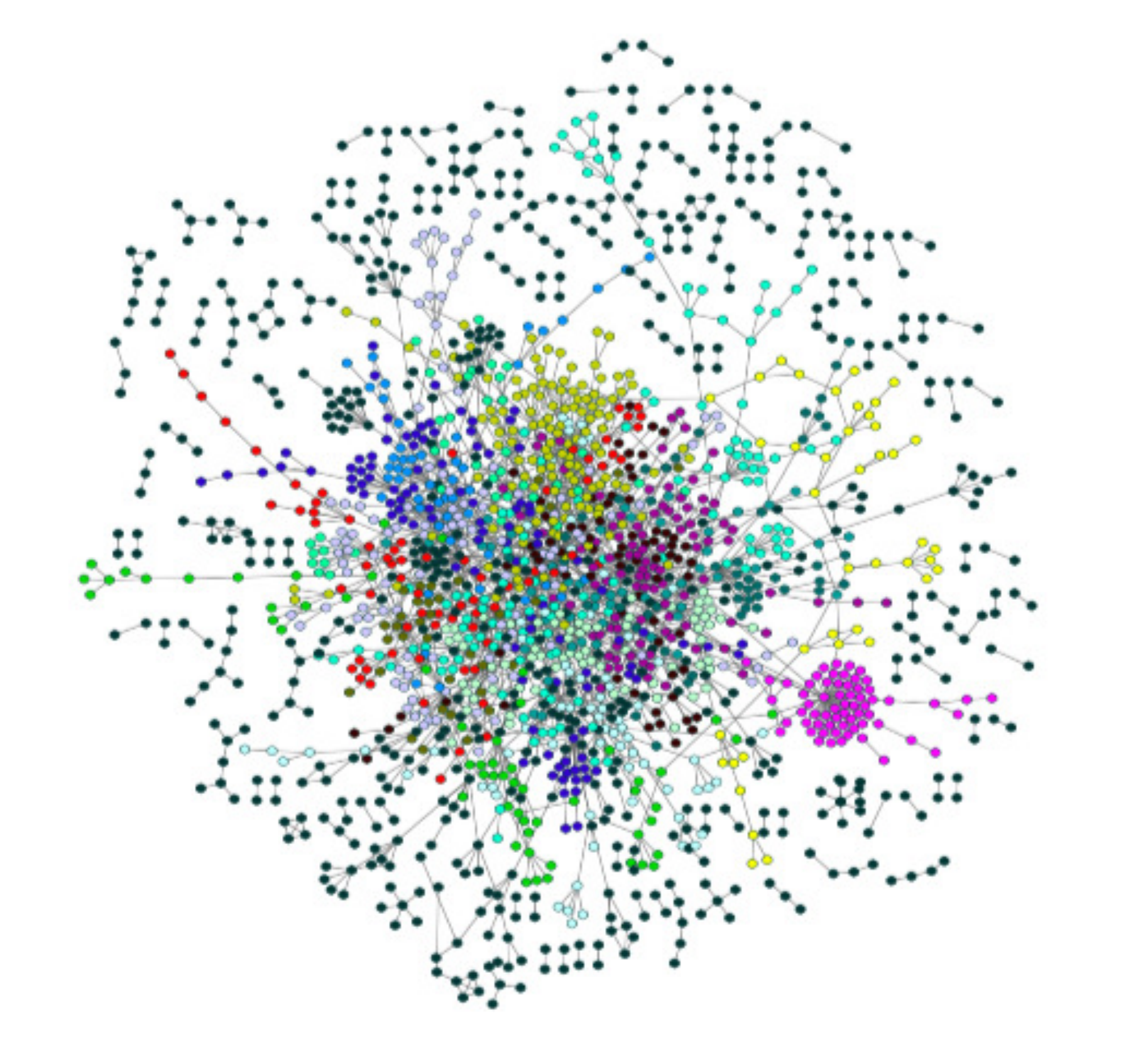}}
  \\
  \subfloat[Nexus 5]{\label{fig:nexus5}\includegraphics[scale=0.4]{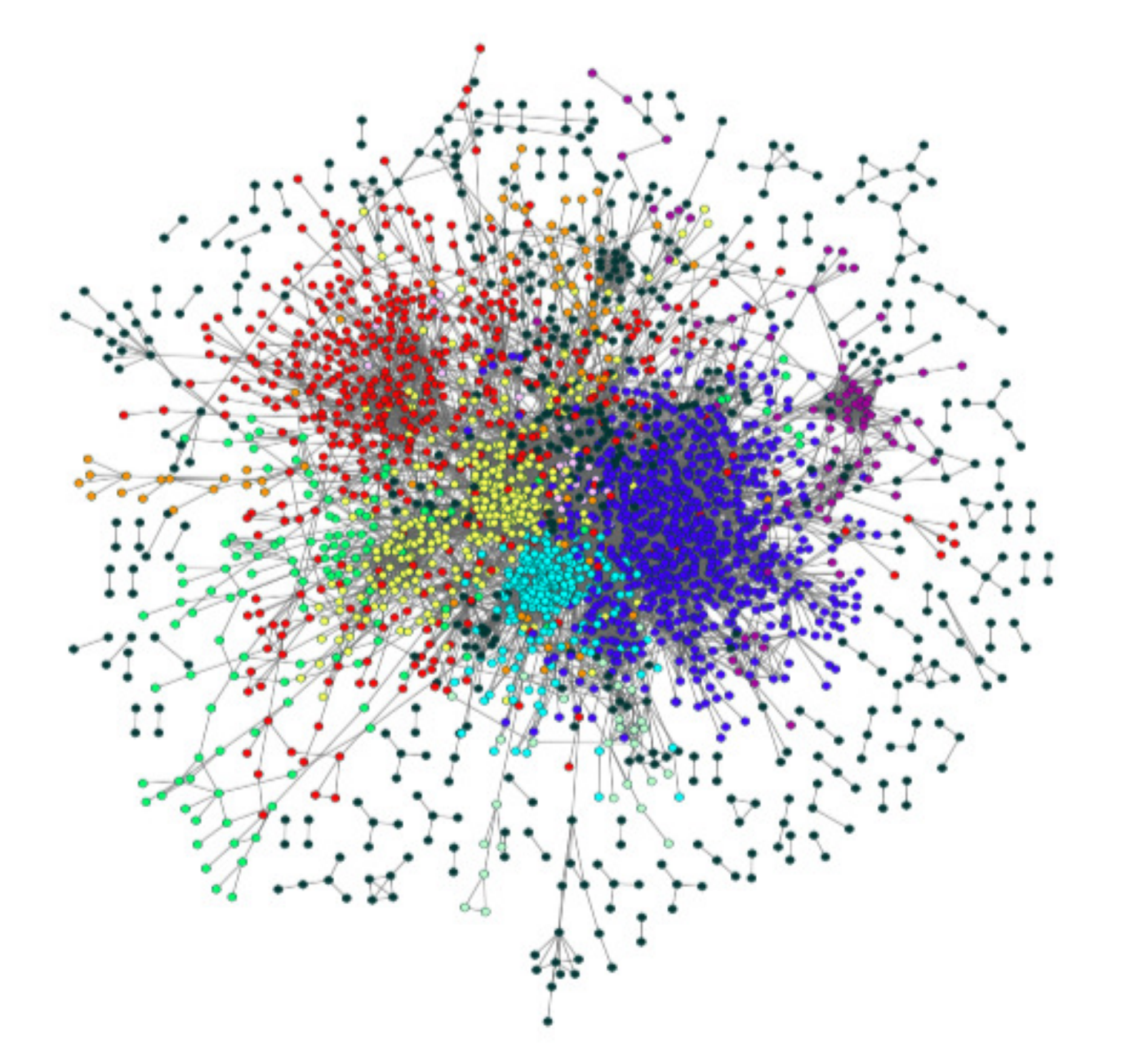}}
  \subfloat[Nexus 15]{\label{fig:nexus15}\includegraphics[scale=0.4]{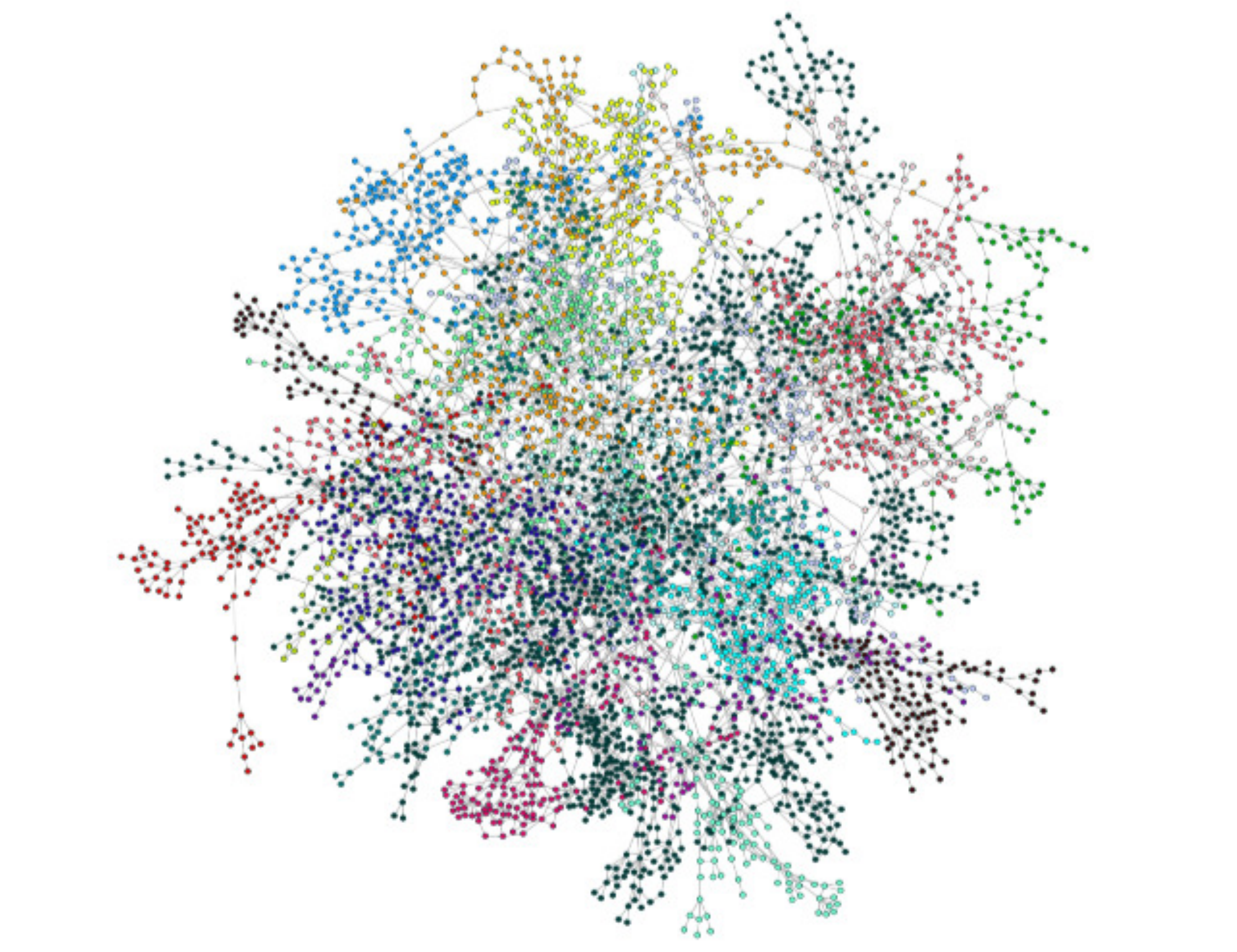}}
  \caption{For each real network (Facebook (a), Barabasi (b), Nexus 5 (c) and Nexus 15 (d)) we show the extracted community found by the proposed method $fast \ greedy$. Only the community with more than the $5\%$ of nodes are displayed.}\label{fig:realnet}
\end{figure}

In order to provide an example of our analysis work-flow, we selected four different publicly available datasets namely two biological (protein-protein interaction networks, $Nexus$ 5 and $Barabasi$), one representing the western states power grid of United States ($Nexus$ 15) and a social dataset ($Facebook$).
Note that $Nexus$ is an online repository of networks, with an $API$ that allow programmatic queries against it, and programmatic data download as well. These functions can be used to query it and download data from it, directly as an $igraph$ graph. The total number of nodes and edges of these four real networks are summarised in Table \ref{NetworkStat} and displayed in Figure \ref{fig:realnet}.

\begin{table}
\caption*{Real Networks summary}
\centering
\begin{tabular}{crrrrc}
\hline
            &Facebook&Nexus 5&Nexus 15&Barabasi& \\
\hline

Nodes & 4039 & 2617 & 4941&1870& \\
Edges & 88234 & 11855 & 6594 & 2240

\\ \hline
\end{tabular}
\caption{Nodes and Edges number for each of the four real analysed datasets } \label{NetworkStat}

\end{table}

\subsubsection*{Nexus 5}
This dataset consists of an undirected protein-protein interaction network in yeast. This data set was compiled by von Mering et al. (see \cite{vonMerin2002}) combining various sources. Only the interactions that have high and medium confidence are included here. 

\subsubsection*{Barabasi}
This dataset consists of the protein-protein interaction network in Saccharomyces cerevisiae described and analysed in \cite{Barabasi2001}. It is derived from combined, non-overlapping data, obtained mostly by systematic two-hybrid analyses. 
Data are available at \url{http://www3.nd.edu/} 
\newline
\url{~networks/resources.htm}.

\subsubsection*{Nexus 15}
This dataset is an undirected, unweighted network representing the topology of the Western States Power Grid of the United States and was compiled by Duncan Watts and Steven Strogatz. 
Data are available at \url{http://cdg.columbia.edu/cdg/datasets}, \cite{Watts1998}.

\subsubsection*{Facebook}
This dataset consists of `circles' (or `friends lists') from Facebook \cite{McAuley2012}.
The authors obtained profile and network data from 10 ego-networks, consisting of 193 circles and 4,039 users. To do so they developed their own Facebook application and conducted a survey of ten users, who were asked to manually identify all the circles to which their friends belonged. On average, users identified 19 circles in their ego-networks, with an average circle size of 22 friends. Examples of such circles include students of common universities, sports teams, relatives, etc. Data are available at \url{http://snap.stan-} \newline \url{ford.edu/data/egonets-Facebook.html}.\\

The application of the overall procedure on the just described real datasets is summarised in Tables \ref{BFreal} and \ref{FADreal} and in Figures \ref{fig:VIfastReal} and \ref{fig:VIlouvainReal}, respectively. 

\subsubsection*{Gaussian Process results}
The application of the Gaussian Processes approach described in Section \ref{sec:GP} to the four real networks is summarised in Table \ref{BFreal}. The resulting BF are very high for either  $fast \ greedy$ or $louvain$ clustering. This gives strong statistical evidence that the four analysed networks have a robust clustering structure hence the recovered community structures are not likely to be random.  

\begin{table}[h]
\centering
\begin{tabular}{ | l | c | c |}
\hline
   Datasets & fast greedy & Louvain \\  \hline
   Barabasi & 284.411 & 243.816 \\ \hline
   Facebook & 297.251 & 361.060 \\ \hline
   Nexus 5   & 340.795 & 431.477 \\ \hline
   Nexus 15  & 503.183 & 495.810 \\ \hline
\end{tabular}
\caption{GP Bayes Factor on the 4 datasets after clustering via $fast \ greedy$ and $louvain$} \label{BFreal}
\end{table}

\subsubsection*{Functional Principal Component testing results} 
Similarly, Table  \ref{FADreal} summarises the application of the Functional $Anderson$-$Darling$ test described in Section \ref{sec:FPC} to the simulated data. As we can see the False Discovery rate adjusted p-values are well lower then the standard significance value $0.05$, after clustering with either clustering  $fast \ greedy$ or $louvain$. This result agree with the previous one leading to the same conclusion that analysed real networks have a robust clustering structure.

\begin{table}[h]
\centering
\begin{tabular}{| l | c | c |}
\hline
   Datsets  & fast greedy & Louvain \\  \hline
   Barabasi & 0.00016  &  0.00302  \\ \hline
   Facebook & 0.00024  &  8e-05 \\ \hline
   Nexus 5   & 0.000765 &  0.00024 \\ \hline
   Nexus 15  & 8e-05    &  8e-05 \\ \hline
\end{tabular}
\caption{FDR adjusted p-values by the FAD test procedure on the 4 datasets after clustering via fast greedy and Louvain.}\label{FADreal}
\end{table}

\subsubsection*{Interval-wise Functional Testing results}
\begin{figure}[h]
\centering
\subfloat[nexus 5]{\includegraphics[width=0.4\textwidth]{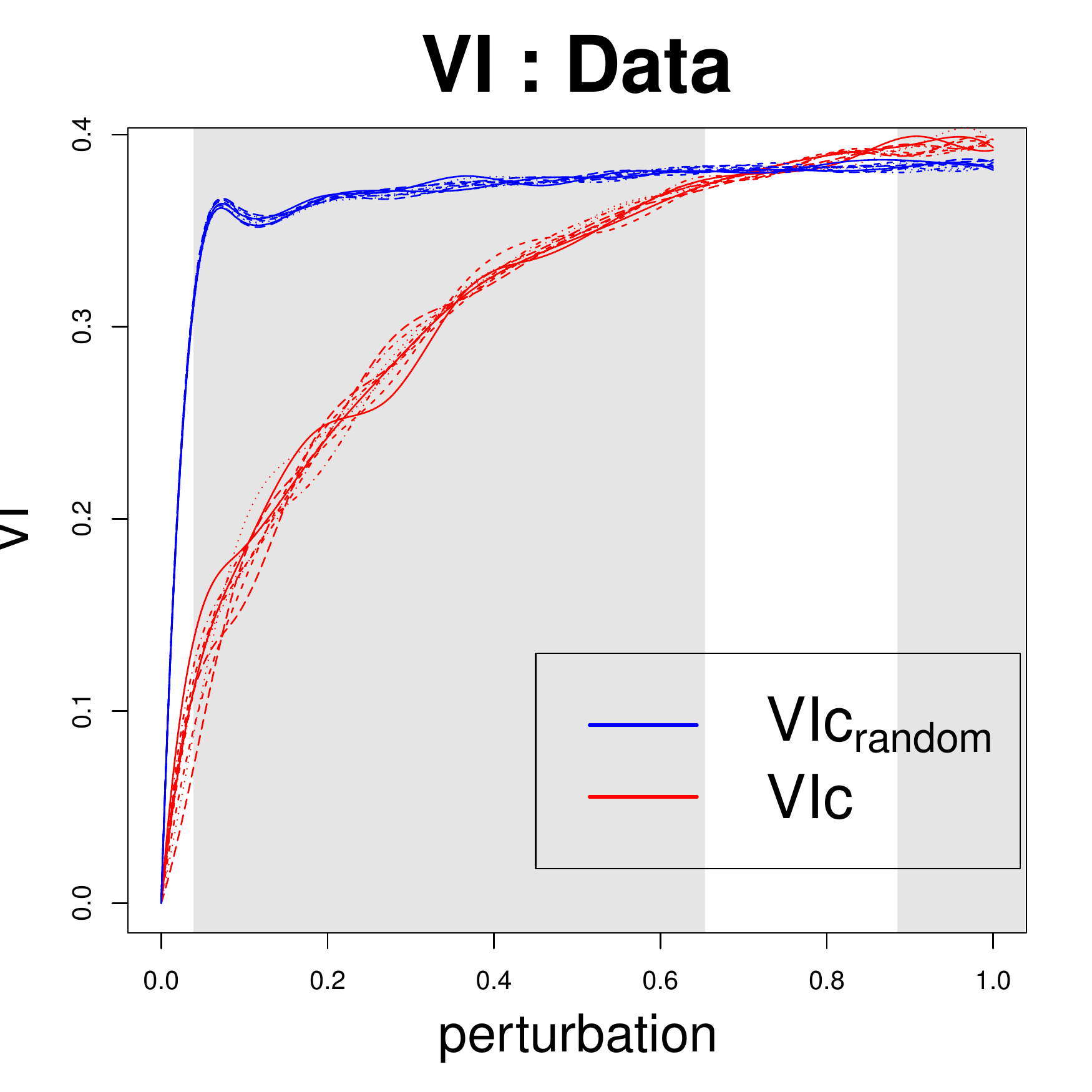}}%
\subfloat[nexus 5]{\includegraphics[width=0.4\textwidth]{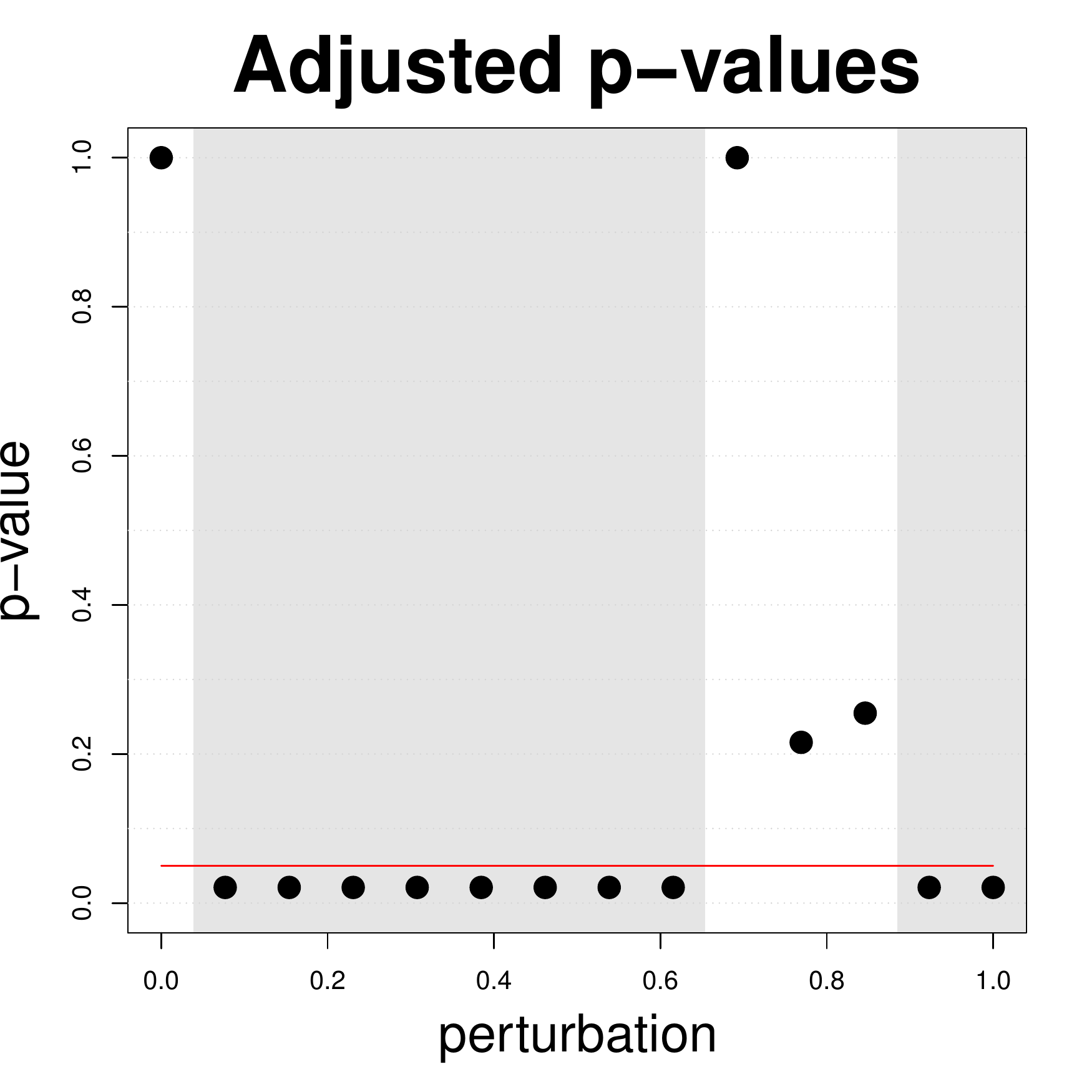}}\\

\subfloat[nexus 15]{\includegraphics[width=0.4\textwidth]{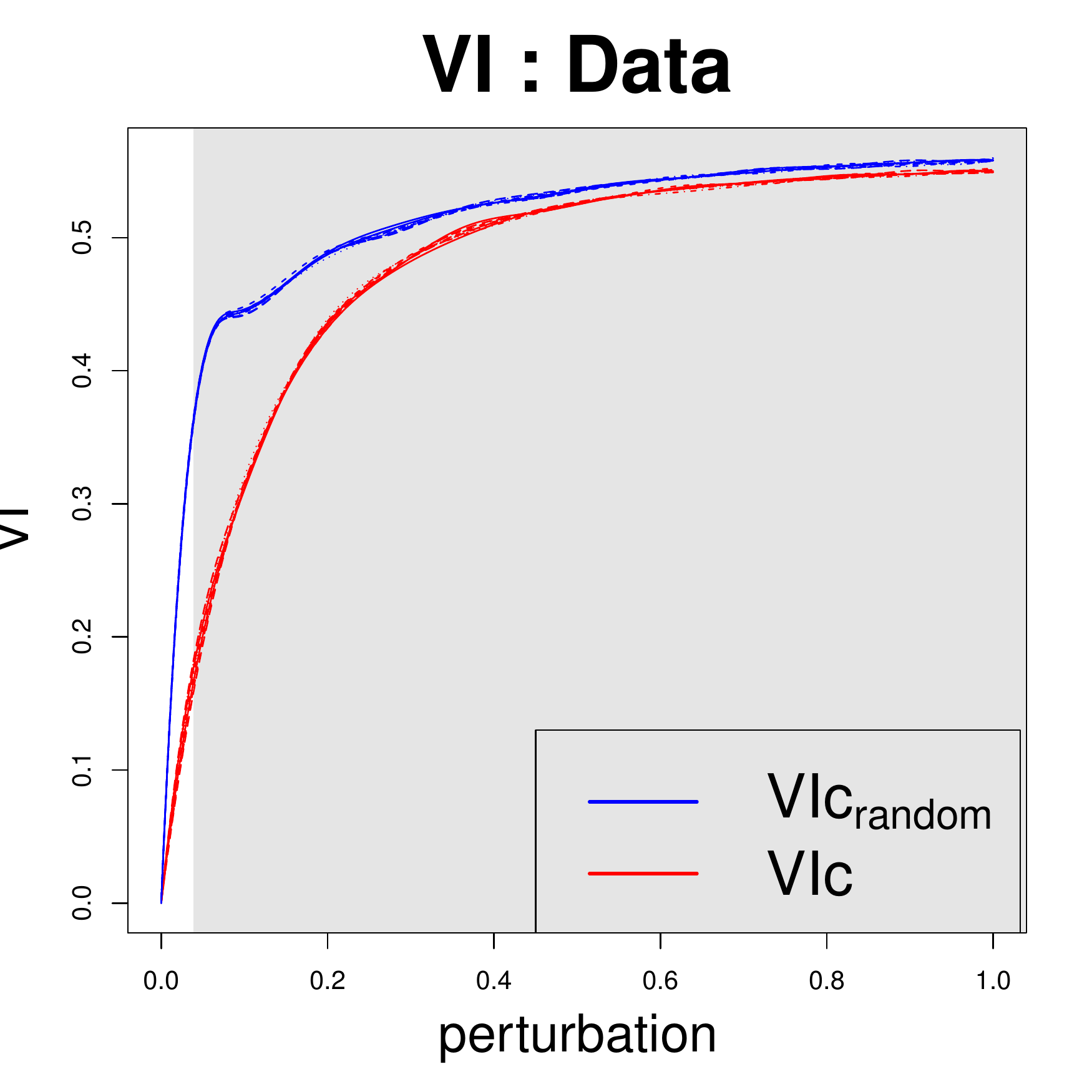}}
\subfloat[nexus 15]{\includegraphics[width=0.4\textwidth]{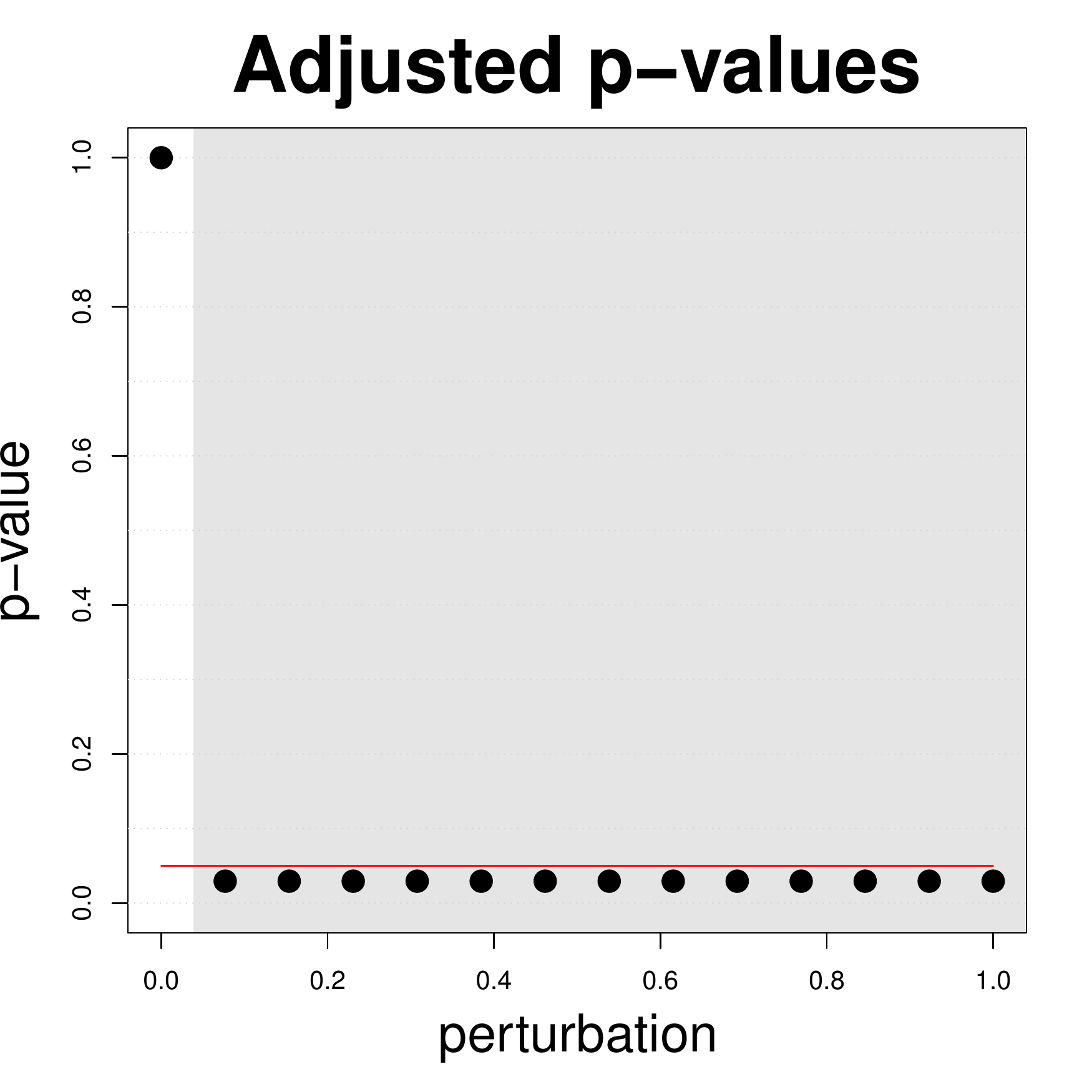}}\\
\caption{VI plots on the clustering obtained via Fast Greedy on real datasets and the corresponding adjusted p-values of the Interval Testing procedure. Horizontal red line corresponds to the critical value 0.05. Light grey areas correspond to p-values below 0.05, dark grey areas correspond to p-values below 0.01.}\label{fig:VIfastReal}
\end{figure}

\begin{figure}[h]
\centering
\subfloat[nexus 5]{\includegraphics[width=0.4\textwidth]{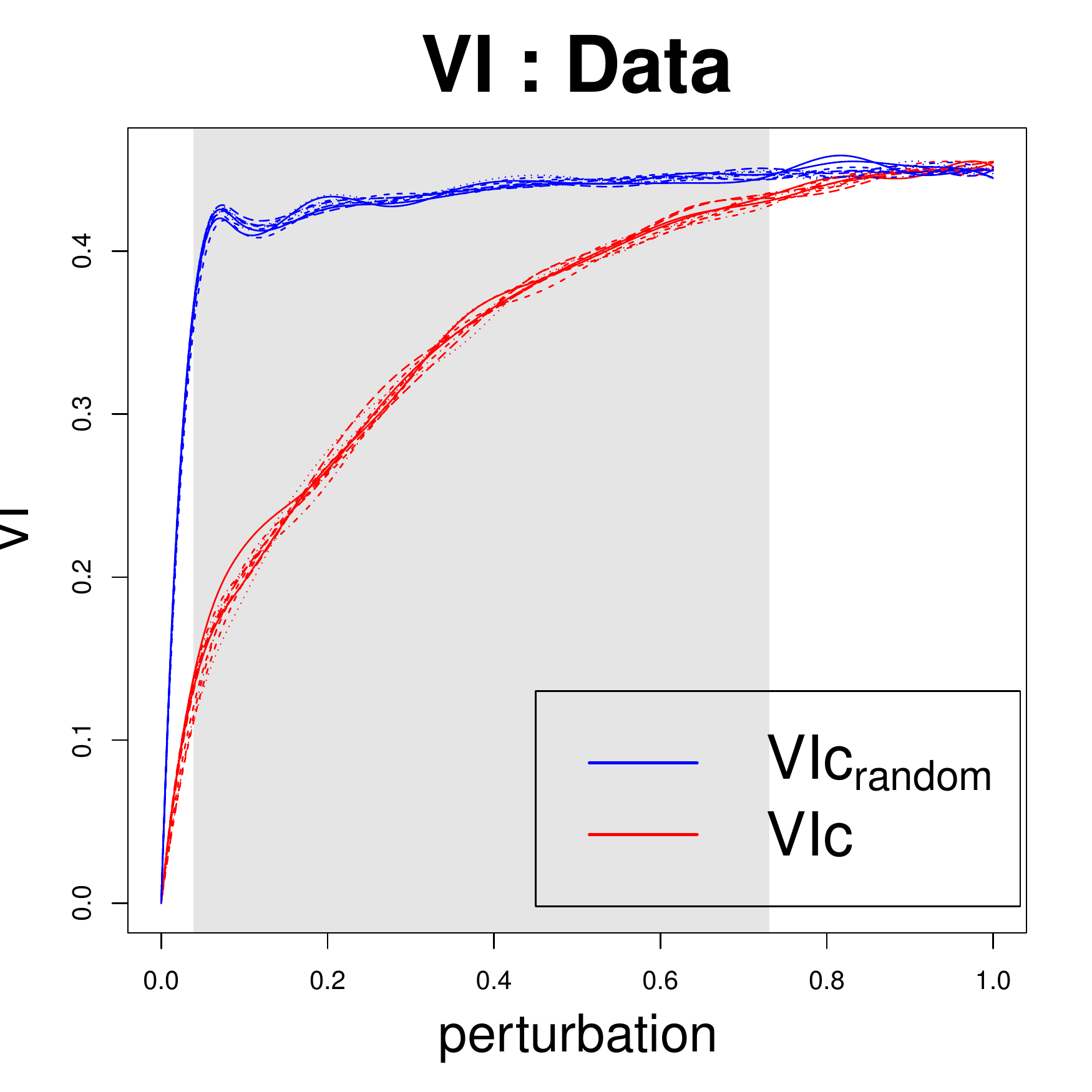}}%
\subfloat[nexus 5]{\includegraphics[width=0.4\textwidth]{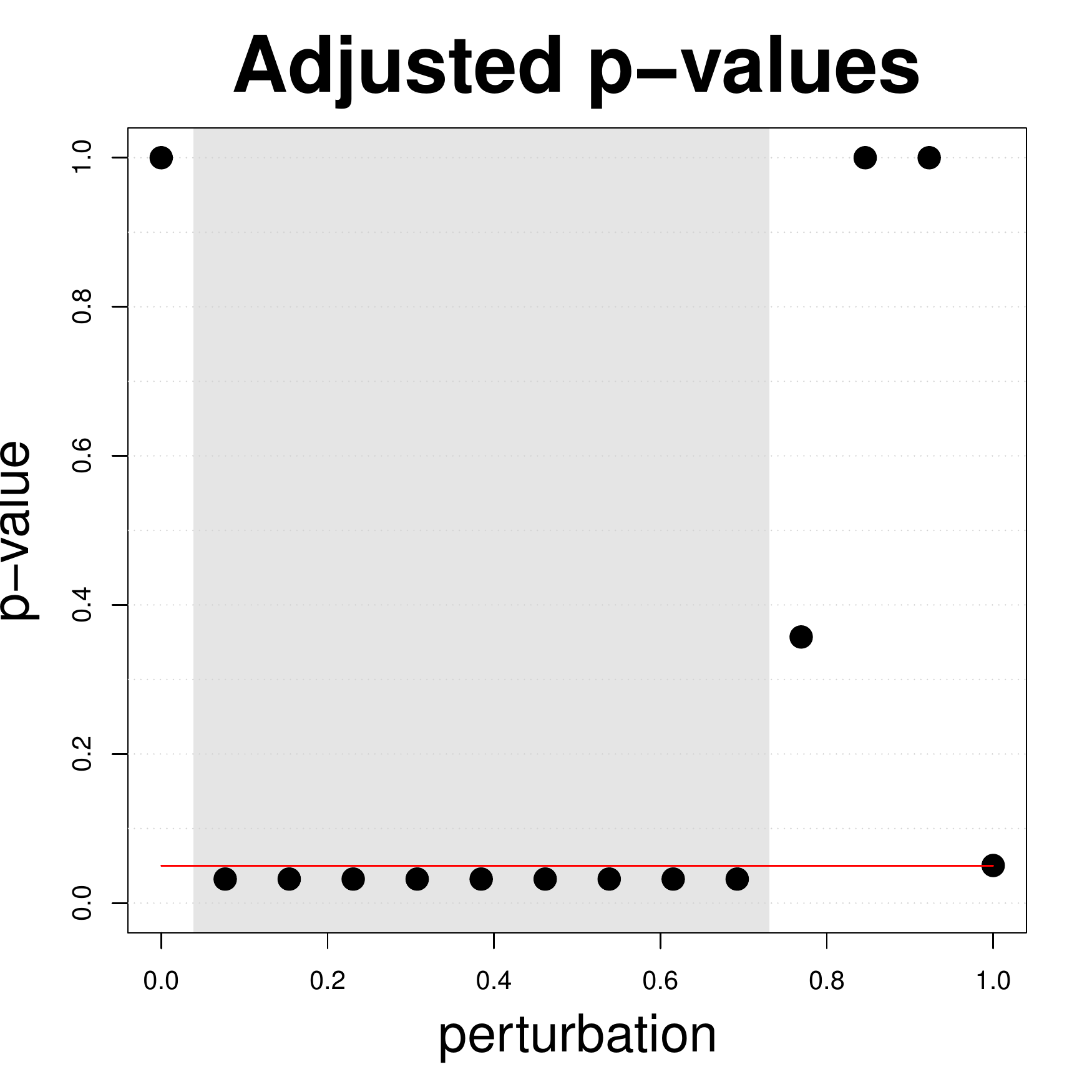}}\\

\subfloat[nexus 15]{\includegraphics[width=0.4\textwidth]{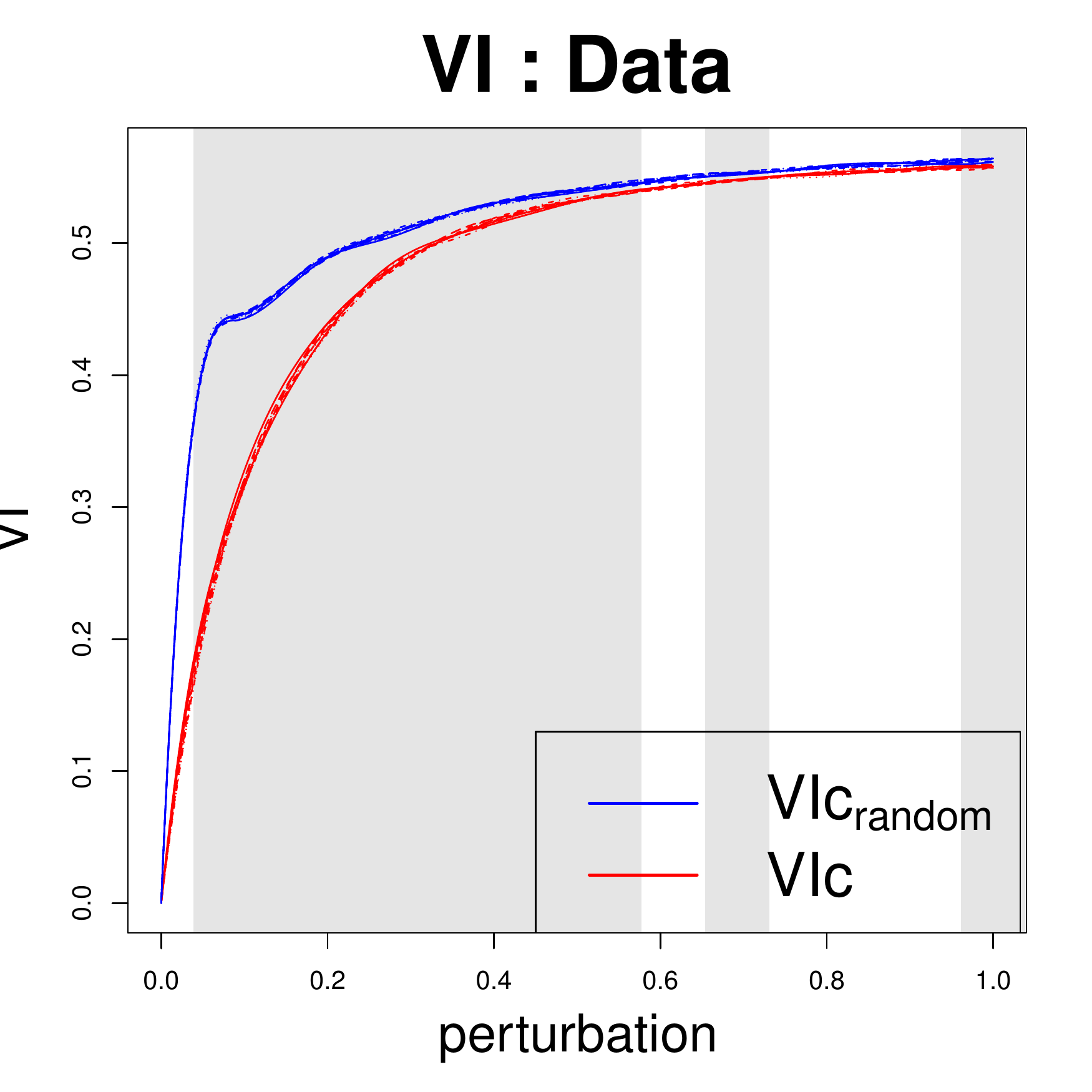}}
\subfloat[nexus 15]{\includegraphics[width=0.4\textwidth]{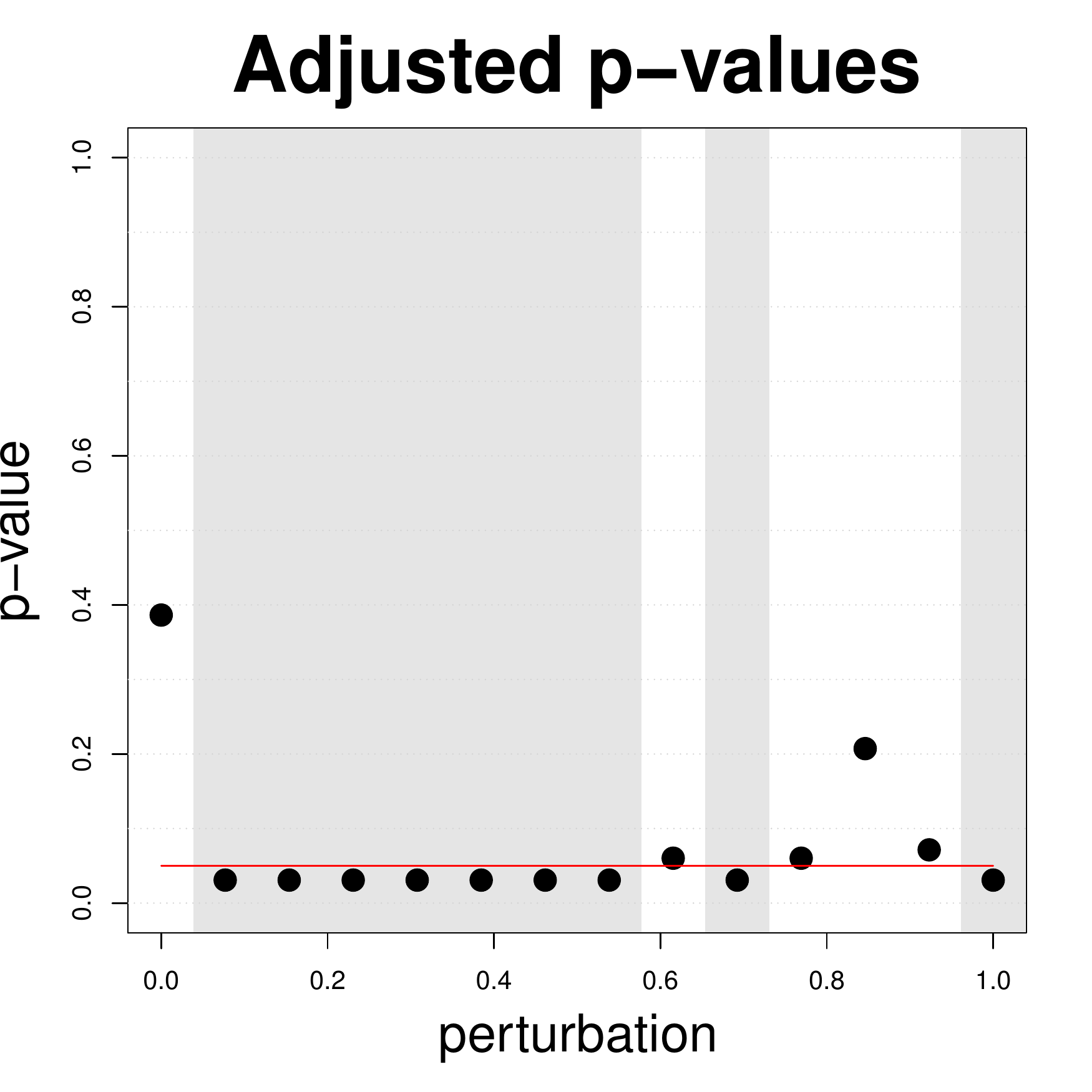}}
\caption{VI plots on the clustering obtained via Louvain on real datasets and the corresponding adjusted p-values of the Interval Testing procedure. Horizontal red line corresponds to the critical value 0.05. Light grey areas correspond to p-values below 0.05, dark grey areas correspond to p-values below 0.01.}\label{fig:VIlouvainReal}
\end{figure}

The application of the Interval-wise Testing procedure described in Section \ref{sec:IFT} to the real datasets after clustering via  $fast \ greedy$ or $louvain$ are depicted respectively in Figures \ref{fig:VIfastReal} and \ref{fig:VIlouvainReal}. 

In each figure, panels $(a),(c),(e),(g)$ show the VI curves for the null model ($VIc_{random}$) and for the actual model ($VIc$). In all the cases the two curves appear to be very close for high perturbation values and depart from each other as perturbation level approaches zero. In panels $(b),(d),(f),(h)$ this is quantified locally by a specific adjusted p-value in each sub-interval. Also in this case significant p-values are falling under the horizontal red line corresponding to the critical value of 0.05. As expected, either using $louvein$ or $fast$ $greedy$ as clustering methods yields to similar results conclusion. As already observed for the synthetic datasets, if we strongly perturb a network ($p\geq 0.5$, i.e. we rewire more than $50\%$ of edges) it approaches a random network, indeed the two $VI$ curves become very close, and the p-value could survive the threshold. 

\begin{figure}[h]
\centering

\subfloat[barabasi]{\includegraphics[width=0.4\textwidth]{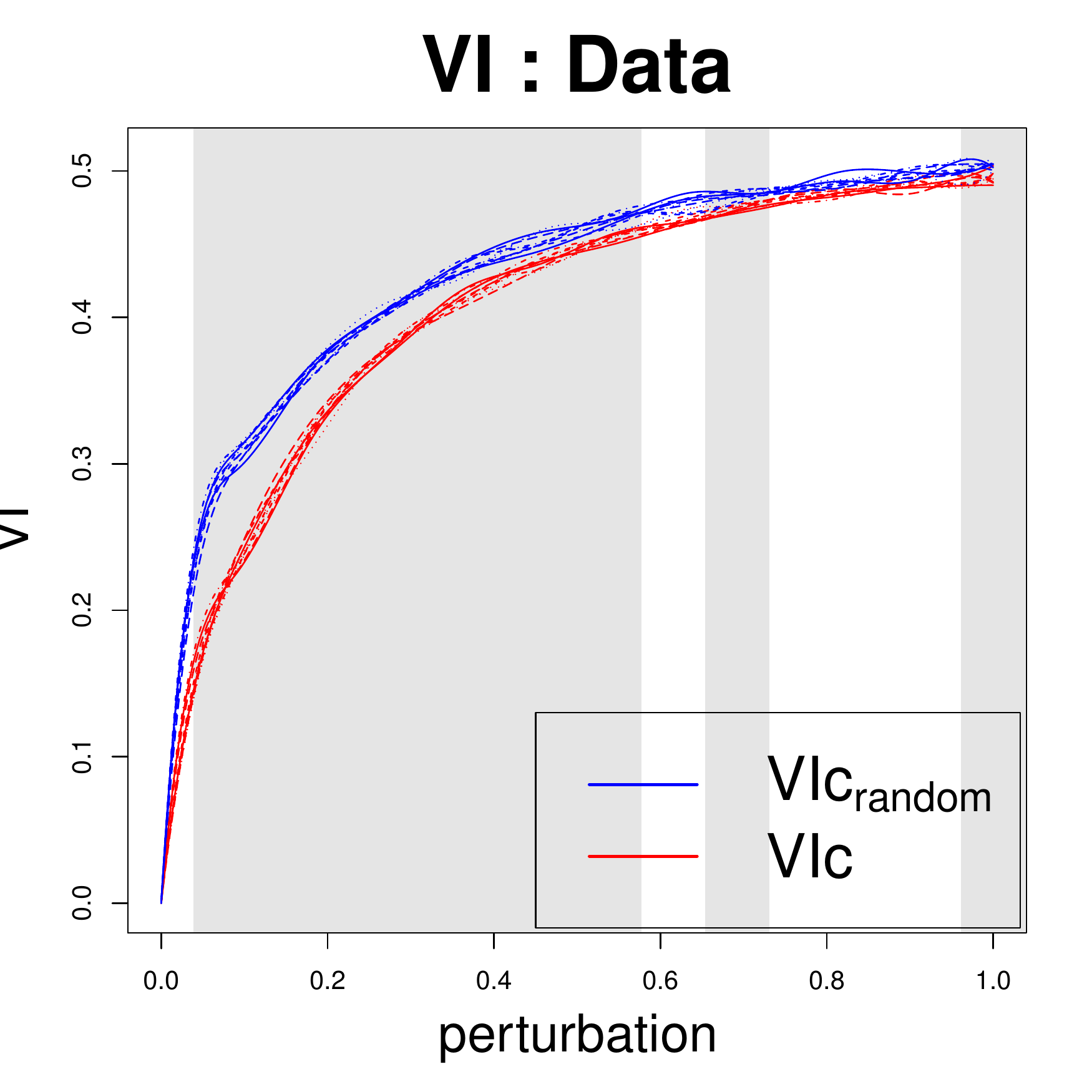}}%
\subfloat[barabasi]{\includegraphics[width=0.4\textwidth]{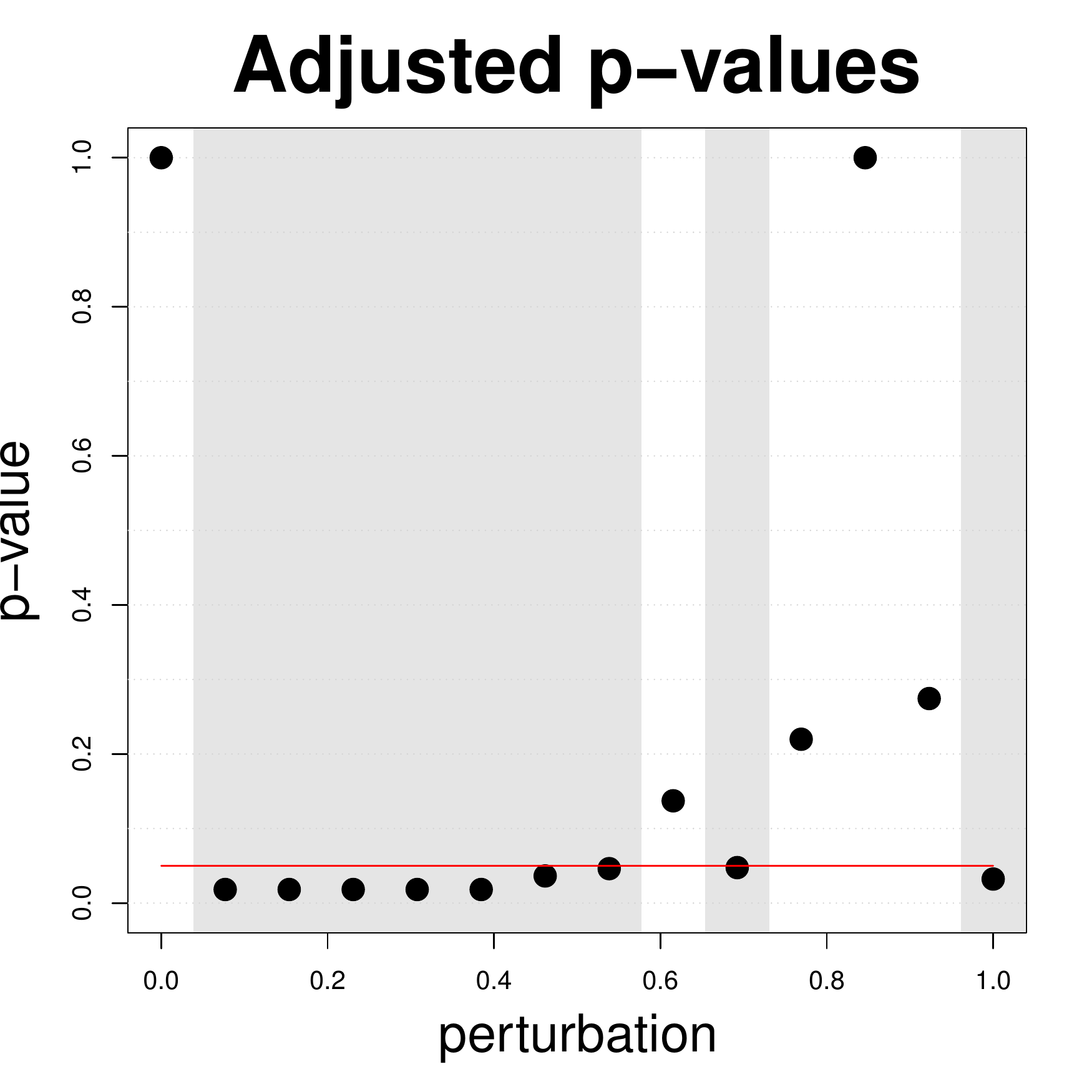}}\\

\subfloat[facebook]{\includegraphics[width=0.4\textwidth]{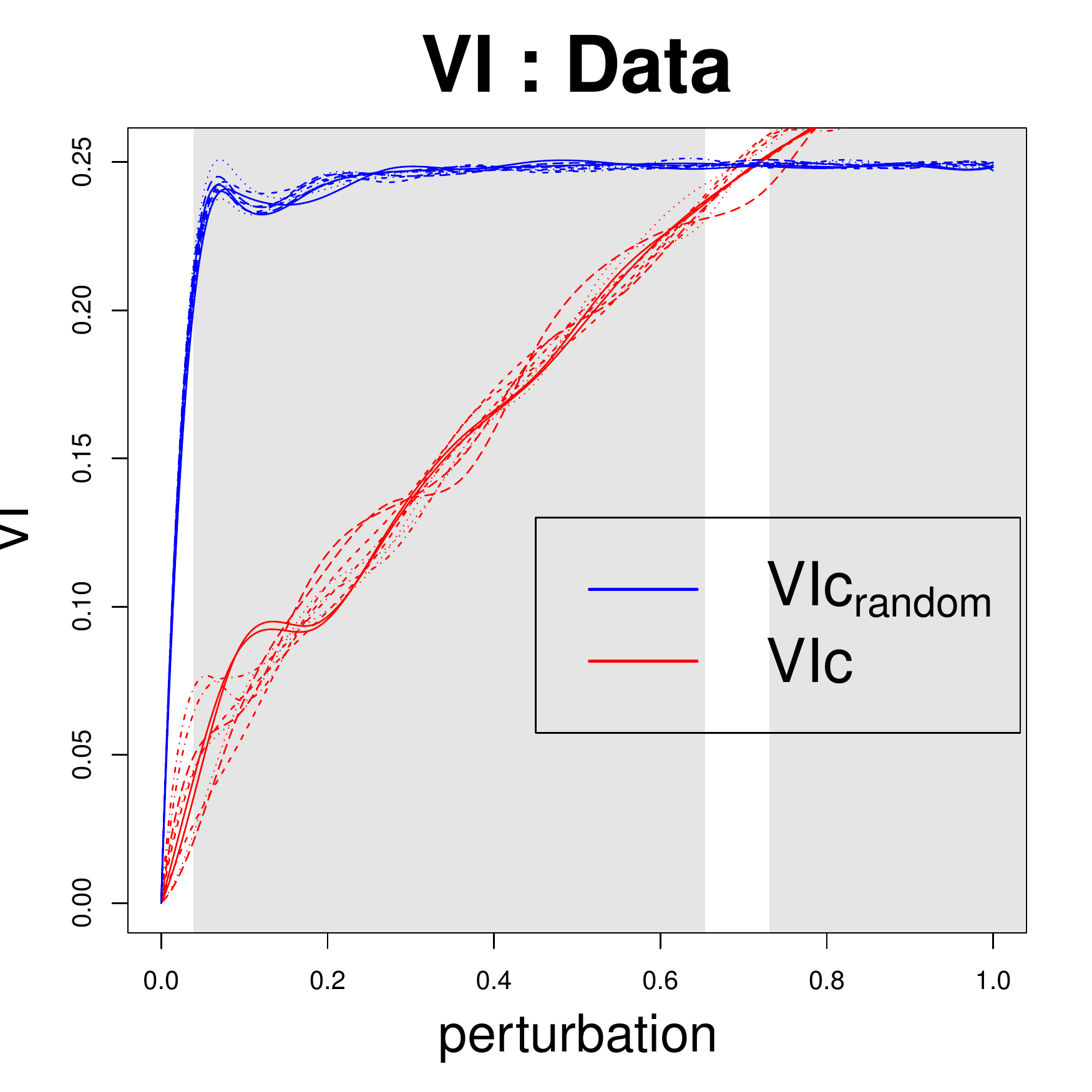}}
\subfloat[facebook]{\includegraphics[width=0.4\textwidth]{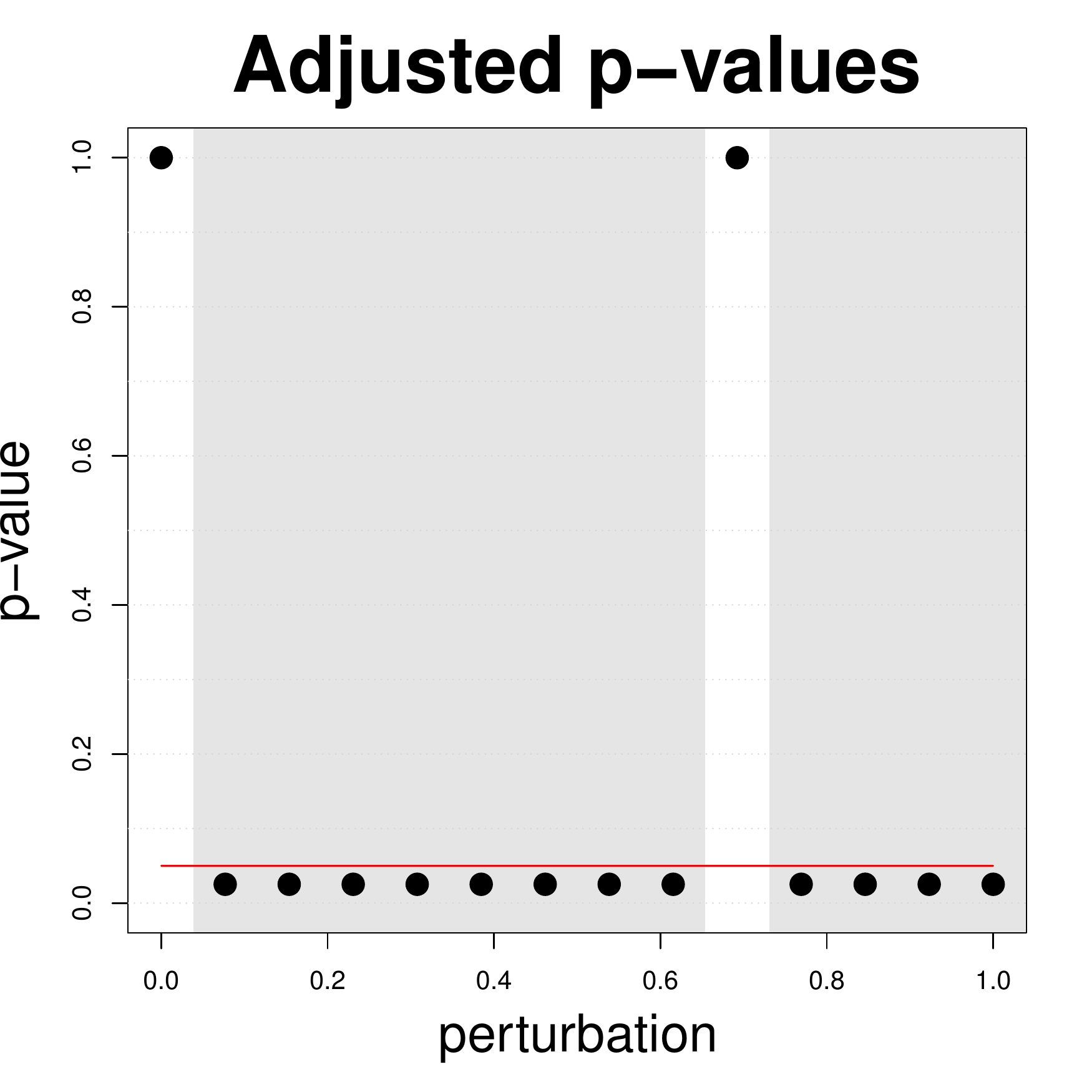}}\\
\end{figure}
\setcounter{subfigure}{4}

\begin{figure}[h]
\centering
\subfloat[barabasi]{\includegraphics[width=0.4\textwidth]{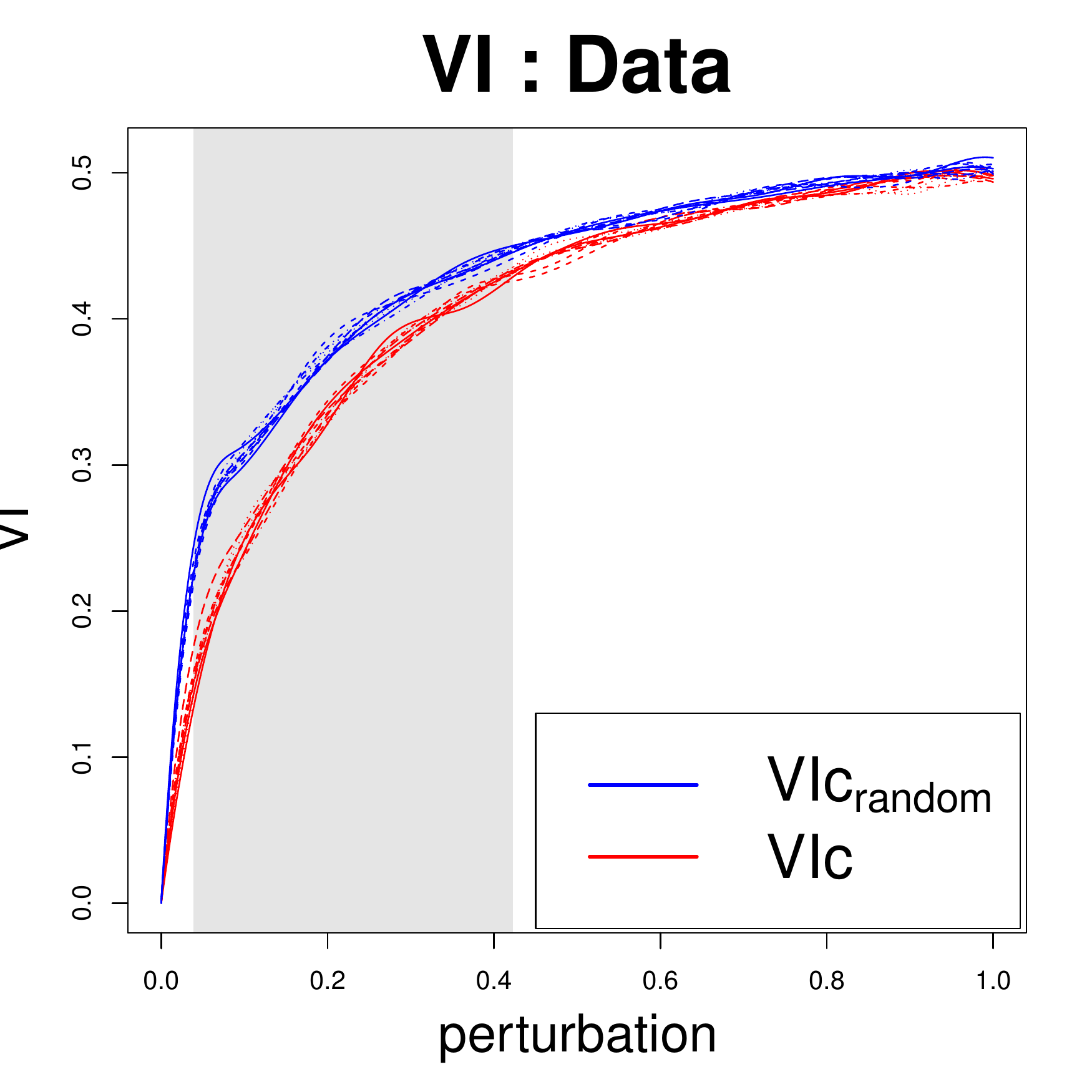}}%
\subfloat[barabasi]{\includegraphics[width=0.4\textwidth]{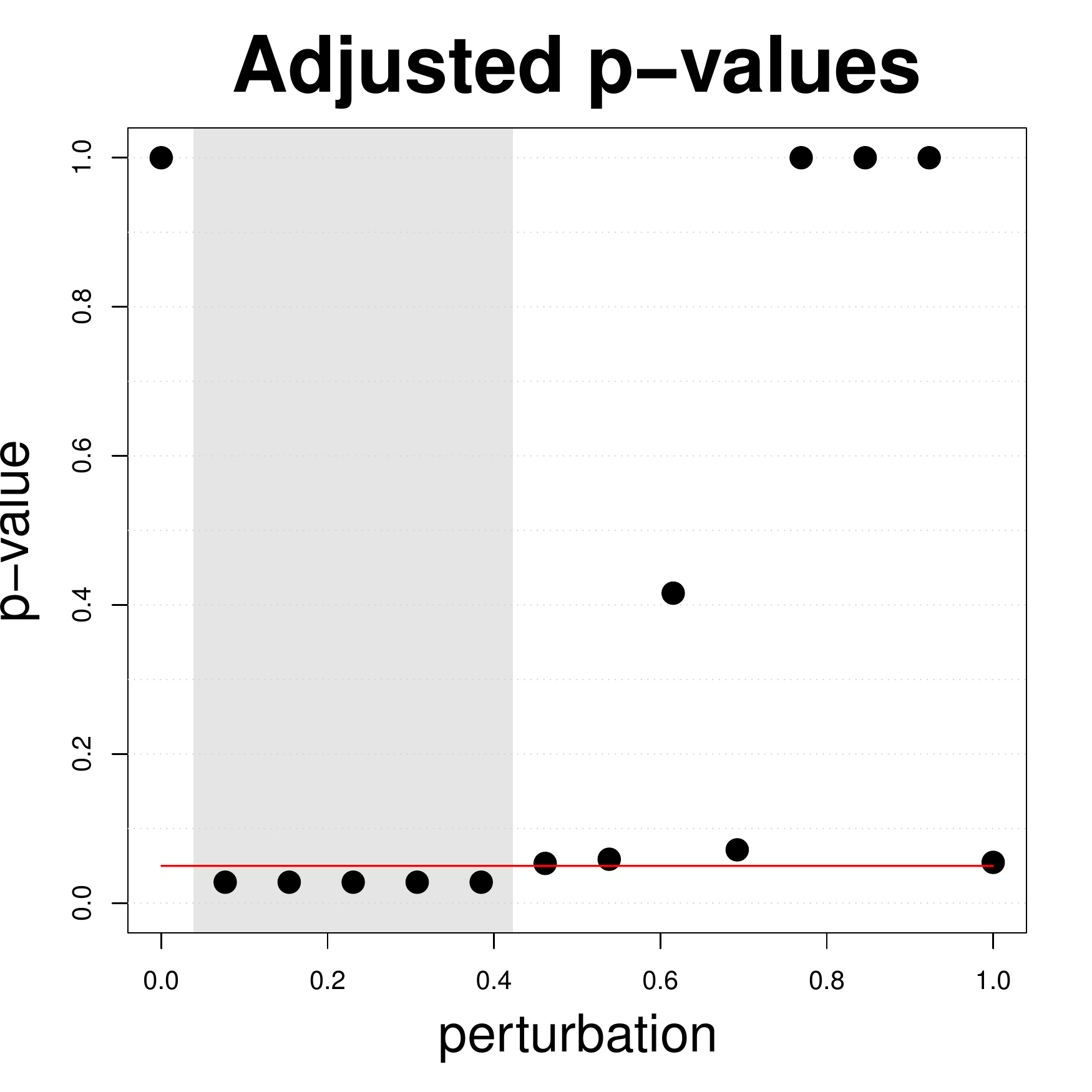}}\\

\subfloat[facebook]{\includegraphics[width=0.4\textwidth]{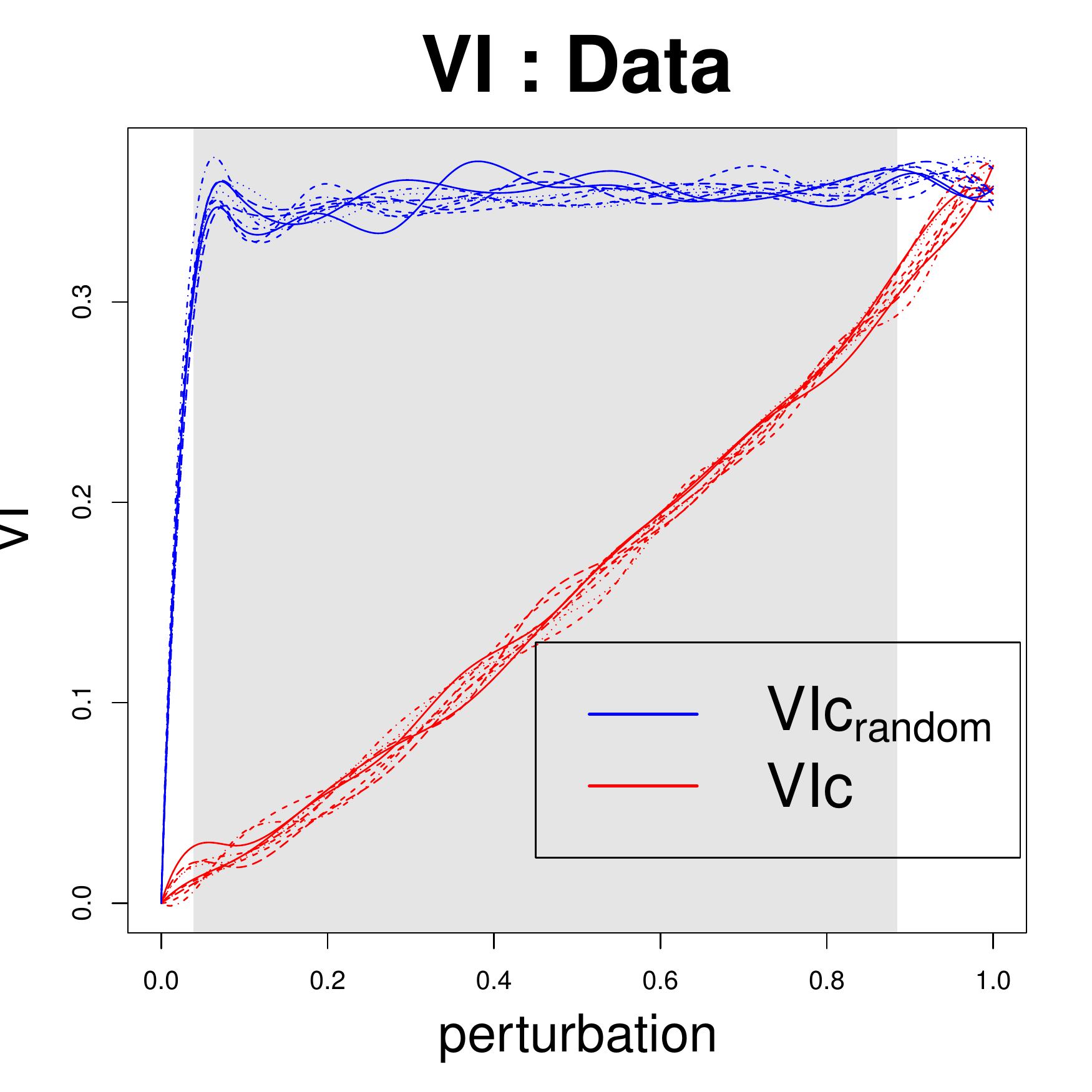}}
\subfloat[facebook]{\includegraphics[width=0.4\textwidth]{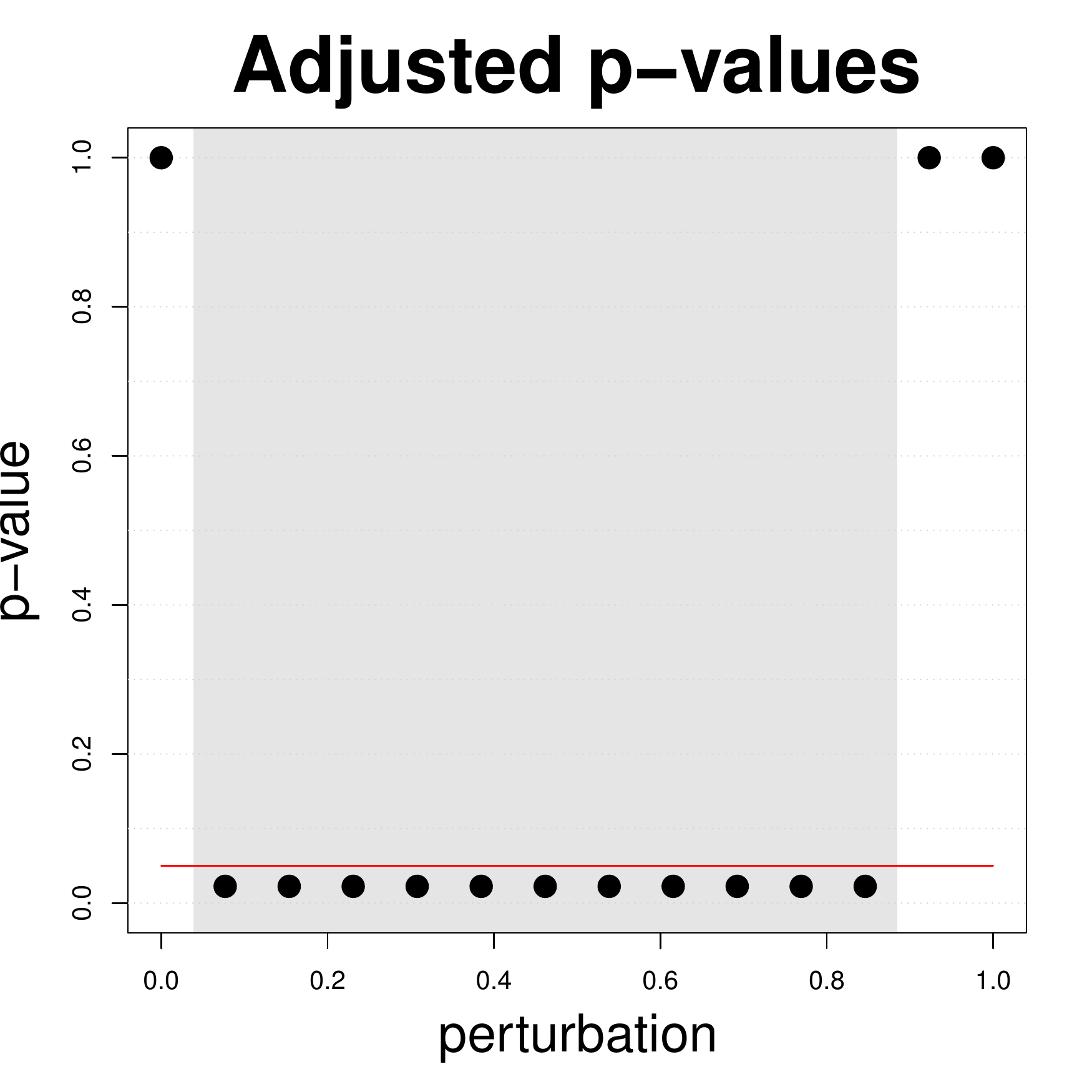}}\\
\end{figure}
\setcounter{subfigure}{4}

\section{Conclusions and Discussions}
\label{Discussion}
In this paper we propose an effective procedure to evaluate the robustness of a clustering.  Given a community detection method and a network of interest, our methodology enables to clearly detect
if the community structure found by some algorithms is statistically significant or is a result of chance, permitting to examine the stability of the partition recovered. As suggested in \cite{Karrer2008}, we specify a perturbation strategy and a null model to build a set of procedures based on VI as stability measure. This enables to build the VI curve as a function of the perturbation percentage and to compare it with the corresponding null model curve in the functional data analysis framework.

We point out that our methodology could also be used to compare different clustering methodologies, indeed given two clusterings on the same network, we could test the agreement between the two recovered partitions via the direct comparison of the corresponding VI curves as defined by our procedure in Section \ref{VI}. For example, all the three procedures we used point out that $louvain$  method is able to recover a non random clustering also at low modularity ($Q\geq 0.2$), the other way around $fast$ $greedy$ needs a more defined structure ($Q\geq 0.4$).

However, it is out of the scope of the present paper the comparative evaluation of different community extraction methods. The two methodologies $louvain$ and $fast$ $greedy$ were indeed only instrumental to the exemplification of our procedure. Both of them were selected at this stage as they both enables for an automatic definition of the optimal number of communities and are based on the optimisation of the modularity, that plays a key role in describing community structures. 

An interesting and straightforward extension of the current paper would be using a different clustering stability measure, for example the $Normalized$ $Mutual$ $Information$ measure proposed in \cite{Danon2005}. This would also lead to a comparison of the performance of different measures for community structure comparison.

\end{document}